\documentclass[twocolumn,showpacs,prb,amsmath,amssymb]{revtex4}

\usepackage{graphicx}

\usepackage{epsfig}

\begin{document}

\title{Disordered mesoscopic systems with interactions: induced
  two--body ensembles and the Hartree--Fock approach}

\author{Y. Alhassid,$^1$ H. A. Weidenm\"uller$^2$ and A. Wobst$^3$}

\affiliation{$^{1}$Center for Theoretical Physics, Sloane Physics
Laboratory, Yale University, New Haven, CT 06520, USA\\
$^2$Max-Planck-Institut f\"ur Kernphysik, D-69029 Heidelberg,
Germany\\$^3$Institut f\"ur Physik, Universit\"at Augsburg,
          86135 Augsburg, Germany\\}

\begin{abstract}

We introduce a generic approach to study interaction effects in 
diffusive or chaotic quantum dots in the Coulomb blockade regime.
The randomness of the single--particle wave functions induces
randomness in the two--body interaction matrix elements. We 
classify the possible induced two--body ensembles, both in the
presence and absence of spin degrees of freedom. The ensembles
depend on the underlying space--time symmetries as well as on
features of the two--body interaction. Confining ourselves to
spinless electrons, we then use the Hartree--Fock (HF)
approximation to calculate HF single--particle energies and HF
wave functions for many realizations of the ensemble. We study
the statistical properties of the resulting one--body HF ensemble
for a fixed number of electrons. In particular, we determine the
statistics of the interaction matrix elements in the HF basis, of
the HF single--particle energies (including the HF gap between
the last occupied and the first empty HF level), and of the HF
single--particle wave functions. We also study the addition of
electrons, and in particular the distribution of the distance
between successive conductance peaks and of the conductance peak
heights. 

\end{abstract}

\maketitle

\section{Introduction}

In this paper, we explore a generic approach to interaction effects in
diffusive and/or chaotic quantum dots in the Coulomb blockade regime.
To set the stage, we describe in this Introduction some previous work,
and discuss the motivation of our own investigation.

Electron transport in quantum dots that are strongly coupled to leads
(open dots) has been successfully described within a single--particle 
framework. Electron--electron interactions affect the dephasing rate
of the electrons at finite temperature, but otherwise electrons in the
vicinity of the Fermi energy can be described as non--interacting 
quasi--particles. The single--particle dynamics in the dot can be
affected by disorder or by the dot's boundaries. In a disordered dot,
the electron moves diffusively. If the disorder is weak, the
statistical fluctuations of the single--electron spectrum and
wave functions can be described by random--matrix theory
(RMT).\cite{rmt98} Similarly, in a ballistic dot with irregular
boundaries, the classical dynamics are mostly chaotic and the
statistical quantal fluctuations also follow RMT. In such diffusive
and/or chaotic open dots, the conductance exhibits ``random'' but
reproducible fluctuations versus, e.g., gate voltage or magnetic
field that can be  successfully described by RMT.\cite{rmp00}

The situation changes as the coupling of the dot to the leads is
reduced and the dot becomes almost isolated. The charge on the dot
becomes quantized, the conductance as a function of gate voltage
displays sharp peaks (the Coulomb blockade resonances), and the
Coulomb interaction between electrons becomes important.\cite{rmp00} 
In this Coulomb blockade regime, the simplest model for a quantum dot
is the constant interaction (CI) model. The interaction is modeled by
the classical charging energy of a system with capacitance $C$. For a
fixed number of electrons $n$ on the dot, the interaction of this
model is constant, and the model essentially reduces to a
single--particle model. While certain statistical properties of the
conductance peak heights can be described within the CI plus RMT
model,\cite{JSA92,alhassid96} other observables, most notably the
distribution of the distance between successive conductance peaks
(a distance known as the peak spacing), deviate significantly from
the predictions of the CI plus RMT
model.\cite{sivan96,simmel97,patel98,luscher01} These deviations have
been understood, at least qualitatively, to follow from residual
interaction effects beyond the charging
energy.\cite{sivan96,blanter97,berkovits98} Such residual interaction effects
are at the center of the present investigation.

In a weakly diffusive and/or chaotic dot, the ``randomness'' of the
single--particle eigenfunctions induces randomness of the interaction
matrix elements. The statistical fluctuations of these interaction
matrix elements (evaluated in the non--interacting basis) were
calculated for a dot with a large Thouless conductance
$g$~\cite{blanter97,mirlin00} and were found to be suppressed. In
the limit $g \to \infty$, only a few interaction terms survive,
leading to the so--called universal
Hamiltonian.\cite{kurland00,aleiner02}  For spinless electrons, the
interaction part of the universal Hamiltonian is composed of just the
charging energy term. This shows that the CI model is appropriate for
spinless electrons and in the limit $g \to \infty$. When the electron
spin is taken into account, there is also an exchange interaction
term, and, in the absence of a time--reversal--symmetry breaking
magnetic field, a Cooper--channel--like term. The exchange interaction
term was found to have important effects on the finite--temperature
peak  spacing distribution.\cite{usaj01,alhassid03}  For finite $g$,
further interaction terms exist and can be calculated within a
systematic expansion of the residual interaction in inverse powers of
$g$. Using a two--body effective screened interaction in the limit of
a small gas constant $r_s$, one can show that for a diffusive dot the
average and the  standard deviation of these residual interaction
terms are at least of the order $\Delta/g$, where $\Delta$ is the
single--particle mean level spacing, while for a chaotic dot the
standard deviation of the residual interaction terms is 
of the order $\Delta \sqrt{\ln g}/g$.\cite{alhassid02a,usaj02} 
For finite values of $g$, such terms affect the ground--state energy
of the dot and various observables. In particular, the position of
the conductance peak height at low temperatures is determined  by the
change of the ground--state energy upon the addition of an electron
to the dot (addition energy). Thus, finite--$g$--corrections to the
universal Hamiltonian affect the peak--spacing
distribution.\cite{alhassid02a,usaj02} 
 
Most studies of interaction effects have been limited to specific
models of a dot. These models include an Anderson model for a
diffusive dot or a billiard model for a ballistic chaotic dot, plus
Coulomb or Hubbard--like (short--range) interactions. For a small
number of electrons, exact numerical solutions are
possible.\cite{sivan96,berkovits98}  For a larger number of electrons,
such models have been studied in the Hartree--Fock (HF)
approximation~\cite{levit99,walker99,cohen99} and in a density
functional approach.\cite{hirose01,hirose02,jiang03}  These models contributed
significantly to our understanding of interaction effects in almost
isolated dots, but they also contain non--generic features.

Our approach to interaction effects in diffusive and/or chaotic,
almost isolated quantum dots aims at a generic understanding of
interaction effects beyond the $g \to \infty$ limit. Our starting
point is a Hamiltonian consisting of a one--body and a two--body part.
The one--body part describes chaos and/or disorder and is, within an
energy interval of width $\approx g \Delta$, described by the
appropriate random--matrix ensemble, i.e., the orthogonal ensemble for
conserved time--reversal symmetry and the unitary ensemble for broken
time--reversal symmetry. The two--body part is given by the Coulomb
interaction. However, the randomness of the single--particle
wave functions induces a randomness of the two--body interaction
matrix elements when the latter are written in terms of the
eigenfunctions of the random one--body part of the Hamiltonian. This
results in a generic ensemble of the two--body interaction that we
derive and use in our studies.

The paper consists of two parts. In the first part, we derive the
statistical properties of the two--body matrix elements (the ``induced
two--body ensemble''). We find that these properties are determined
both by the underlying space--time symmetries of the system and by
features of the two--body interaction. We classify the possible forms
of the first two cumulants of the interaction matrix elements. This is
accomplished by writing these cumulants in a covariant form, ensuring
invariance under a change of the single--particle basis. The first
moments (average values) reproduce the universal Hamiltonian in the
limit $g \to \infty$. The second cumulants are characterized by a
constant $u^2$ ($u^2\propto \Delta^2/g^2$ for a diffusive dot and 
$u^2 \propto \Delta^2 \ln g / g^2$ for a chaotic dot). The invariance 
classes of the second cumulants are determined by the symmetry
properties of the interaction matrix elements under permutations of
the single--particle orbitals. These symmetries are different for,
e.g., a contact interaction or a short-- but finite--range
interaction. We note that the two--body part of one class of the
ensembles which we construct, coincides with the two--body embedded
Gaussian ensembles.\cite{mon75}  These ensembles were introduced in
the nuclear shell model to study the validity of RMT for the
statistical description of nuclear spectra.\cite{french70,bohigas71}
The spectral properties of these embedded ensembles were recently
analyzed using an eigenvector expansion of the second
moments.\cite{benet01,benet03} Combined with a random--matrix
one--body part, such a two--body ensemble was used to study
interaction effects~\cite{alhassid00,alhassid02} and ground-state
 magnetization~\cite{jacquod00} in quantum dots with a small number of
electrons. The spin structure of a system with a spin--conserving random 
interaction was studied in Ref.~\onlinecite{kaplan01}.

In the presence of spin degrees of freedom, one needs to use
non--antisymmetrized two--body matrix elements, and our classification
of the second cumulants is generally done for such elements. The
spinless case requires only antisymmetrized matrix elements. Their
second cumulants are obtained directly from the cumulants of the
non--antisymmetrized elements. The two--body ensembles that correspond
to a contact interaction exist only in the presence of spin.

In the second part of the paper, we use the Hartree--Fock (HF)
approximation to work out interaction effects for some of the
ensembles introduced in the first part. It is difficult to study the
properties of ensembles involving a two--body interaction. Exact
numerical solutions can be obtained only for dots with a small number
of electrons. For a larger number of electrons, it is necessary to
make approximations. The HF approximation serves this purpose. It has
the advantage of being based on a single--particle picture (optimized
to take into account interaction effects), and one can still use
concepts familiar from the independent--particle approach. While the
HF approximation has been studied in the context of specific models
for an interacting dot,\cite{levit99,walker99,cohen99}  the present
work is free of any model--dependent effects and constitutes a generic
statistical HF approach.

Our HF studies are limited to spinless electrons. We investigate, in
particular, the statistical properties of the interaction matrix
elements in the basis of HF single--particle eigenfunctions, and of
the HF single--particle eigenvalues and eigenfunctions. The interest
in such work derives, among others, from generic studies of quantum
dots with a large number of electrons. These often use a HF
single--particle basis (rather than a basis defined in terms of
non--interacting electrons).\cite{alhassid02a}  It is then often
assumed that the statistical properties of the matrix elements in the
HF basis are similar to those of the non--interacting
basis.\cite{blanter97}  Our work permits us to test such assumptions,
whereas previous studies (done in the framework of specific models of
a dot with interactions) have focused on the spectral statistics of
the HF levels only. We find that, in the limit of small interaction
strength $u$, the average interaction matrix elements in the HF basis
acquire a $u^2$ correction (this explains the increase of the average
HF gap with $u$), while the second cumulants remain very close to
their input values. As for the spectral statistics, our generic
studies confirm the conclusions of previous studies. The
nearest--neighbor spacings of filled and empty HF levels follow the
Wigner distribution of RMT, while the distribution of the HF gap
deviates from this distribution. An open problem has been whether
correlations of filled and empty levels follow or deviate from RMT.
Our studies of an appropriate spacing correlator indicate that the gap
weakens the correlations between spacings on its opposite sides. On
the other hand, this spacing correlator is found to follow RMT within
the filled or empty levels alone, adding further support to the
conjecture that the filled and empty levels separately satisfy RMT
statistics. We also show that the HF ensemble satisfies orthogonal
(unitary) invariance, and therefore its wave function statistics is
that of RMT.

We use the statistical HF approach to study the statistics of
observables, and in particular of the peak--spacing and peak--height
distributions. The results can be easily interpreted in Koopmans'
limit,\cite{koopmans34} which assumes that the HF wave functions
remain unchanged upon the addition (or removal) of an electron. 
Koopmans' limit has been investigated in previous HF
studies,\cite{levit99,walker99,cohen99}  but the present approach
again has the advantage of being generic. It allows us to define
the limitations of Koopmans' approach, and to go beyond that limit.
In particular, we find that the peak--spacing distribution can be
described as a convolution of a Wigner distribution (with an
appropriate mean level spacing) and a Gaussian whose width is
approximately the standard deviation of a diagonal interaction
matrix element. On the other hand, we find that the peak--height
distribution is not affected by the residual interaction.

The outline of this paper is as follows: In Section \ref{ran}, we
discuss the induced two--body ensembles. We classify the possible
ensembles for the orthogonal and unitary symmetries in
Sections~\ref{ortho} and \ref{unitary}, respectively, both in the
presence and absence of the spin degrees of freedom. The possible
ensembles are related to different types of two--body interactions, as
characterized by the symmetries of their matrix elements under
permutations of the orbitals. In Section~\ref{HF} we discuss the HF
approach to disordered (or chaotic) systems in the framework of the
many--body ensembles. A perturbative approach that provides a closed
expression for the average HF matrix elements for a fixed
single--particle spectrum is described in Section~\ref{mom}. Numerical
studies of both the first and second moments of the interaction matrix
elements in the HF basis are presented in Section~\ref{HF-elements}.
The statistics of the HF single--particle energies including filled
levels, empty levels and the gap are studied in Section~\ref{HF-spect},
while the statistics of the HF wave function components are studied in
Section \ref{HF-wavefunctions}. Finally in Section~\ref{addition} we
discuss the addition of electrons in the HF approach to disordered
systems, and in particular the peak--spacing and peak--height
distributions.

\section{Disordered systems with interactions}
\label{ran}

We consider a disordered system of interacting electrons, e.g., a
diffusive quantum dot. For simplicity, we first discuss the case of
spinless electrons. The spin degrees of freedom are introduced
later in this Section. For a given realization of disorder, we
describe the system by a Hamiltonian with a one--body part and a
two--body interaction,
\begin{equation}
\label{ensemble}
H = \sum_{i,j} h^{(0)}_{ij}a^\dagger_i a_j + {1\over 2} \sum_{ijkl}
v_{ij;kl}a^\dagger_i  a^\dagger_j a_l a_k \;.
\end{equation}
The $m$ orthonormal single--particle states $|i\rangle = a^\dagger_i
|-\rangle$ ($i=1,\ldots,m$) with $|-\rangle$ the vacuum state, form a
fixed basis with $m \gg 1$. The one--body part $h^{(0)}_{ij}$ of $H$
stands for kinetic energy plus disorder potential, while $v_{ij;kl}$
represents the (screened) Coulomb interaction. The $v_{ij;kl} \equiv
\langle ij|v|kl \rangle$ are non--antisymmetrized matrix elements of
the two--body interaction ($|ij\rangle$ is a product state of particle
1 in state $|i\rangle$ and particle 2 in state $|j\rangle$). For
spinless fermions, we can rewrite the interaction in terms of
antisymmetrized matrix elements $v^A_{ij;kl} \equiv v_{ij;kl} -
v_{ij;lk}$ as
\begin{equation}
\label{ensemble-a}
H = \sum_{i,j} h^{(0)}_{ij}a^\dagger_i a_j + {1\over 4} \sum_{ijkl}
v^A_{ij;kl}a^\dagger_i  a^\dagger_j a_l a_k \;.
\end{equation}

Disorder is taken into account by postulating that $h^{(0)}_{ij}$ is
random. The way this randomness is modeled turns out to be rather 
immaterial.
Indeed, we recall that for weak disorder and in the long--wavelength
limit, all models for non--interacting disordered systems yield the
same supersymmetric field theory, the non--linear supersymmetric sigma
model.\cite{efetov83}  This field theory contains one essential
parameter, the dimensionless Thouless conductance $g$. We will work
out properties of the system described by the Hamiltonian of
Eq.~(\ref{ensemble-a}) in terms of an expansion in inverse powers
of $g$. We will carry the expansion up to and including terms
of order $1/g^2$.

The non--linear sigma model shows that to zeroth order in $1/g$,
$h^{(0)}_{ij}$ is equivalent to one of the canonical ensembles of
RMT, the Gaussian orthogonal (GOE), unitary (GUE)
or symplectic (GSE) ensemble of random matrices, depending upon the
symmetry of the problem.\cite{efetov83,rmt98}  Corrections calculated
by using the expansion in inverse powers of $g$ yield deviations from
predictions of RMT but do not violate the
fundamental symmetries of the problem (invariance under orthogonal,
unitary or symplectic transformations of the basis of single--particle
states). Indeed, the derivation of the non--linear sigma model is
based upon these symmetries, which are thus deeply embedded in the
model and are valid beyond canonical RMT. This fact
will be important as we will heavily rely on these symmetries to
determine the effective form of the Hamiltonian $H$. We will focus
on the cases of orthogonal and unitary symmetry and will not
consider symplectic ensembles in the sequel.

The randomness of $h^{(0)}_{ij}$ induces a degree of randomness into
the two--body matrix elements $v_{ij;kl}$. To see this we recall that
every observable ${\cal O}$ is defined as an average over the ensemble
$h^{(0)}_{ij}$. Because of the invariance of that ensemble under
orthogonal (or unitary or symplectic) transformations of the 
single--particle basis, ${\cal O}$ must likewise be invariant under
all such transformations. Therefore, ${\cal O}$ can depend on the
two--body interaction $v_{ij;kl}$ only through invariants constructed
from the matrix elements $v_{ij;kl}$. Other properties of the Coulomb
interaction that are not encapsulated in the invariants are not
relevant. On a formal level, this fact allows us to define equivalence
classes of two--body interactions. Members of the same class share the
same set of invariants although the interactions may differ in other
properties. All two--body interactions which belong to the same class
as the Coulomb interaction will give rise to identical average
properties of the ensemble~(\ref{ensemble}). These members may be said
to form an induced ensemble of two--body interactions. The word
``induced'' indicates that the ensemble owes its existence to the
randomness of the one--body part $h^{(0)}_{ij}$. Thus, we deal
effectively with an induced two--body random ensemble although the
Coulomb interaction {\it per se} is fixed and not random at all. We
aim at reformulating the ensemble of Hamiltonians $H$ in
Eq.~(\ref{ensemble}) in such a way that only the relevant features
(i.e., the invariants) of $v_{ij;kl}$ are kept. In doing so, we do not
follow strictly the line of argument just given because we perform
separately the averages over the eigenfunctions and eigenvalues of
$h^{(0)}_{ij}$. The induced two--body ensemble will result from the
average over the eigenfunctions.

We accordingly transform $H$ in Eq.~(\ref{ensemble-a}) to the basis
$|\alpha\rangle$ of single--particle eigenstates of $h^{(0)}$. This
yields
\begin{equation}
\label{eigen-ensemble}
H = \sum_{\alpha} \epsilon_{\alpha} a^\dagger_\alpha a_\alpha + {1
  \over 4} \sum_{\alpha \beta \gamma \delta} v^A_{\alpha \beta; \gamma
  \delta} a^\dagger_\alpha a^\dagger_\beta  a_\delta a_\gamma\;.
\end{equation}
In an interval of length $\sim g \Delta$ (where $\Delta$ is the average
single--particle level spacing), the eigenvalues $\epsilon_\alpha$
obey canonical random--matrix statistics. Because of the randomness of
the single--particle eigenfunctions, the elements $v^A_{\alpha \beta;
\gamma \delta}$ are also random. Moreover, the $\epsilon_\alpha$'s are
uncorrelated with the eigenfunctions and, thus, with the $v_{\alpha
\beta; \gamma \delta}$'s. This last property holds strictly if
$h^{(0)}$ stands for the GOE or GUE, i.e., in the limit $g \to
\infty$. For a diffusive system, it holds up to and including terms of
order $1/g^2$.

The number of single--particle levels $\alpha$ is usually chosen to be
finite. Thus, the interaction in Eq.~(\ref{eigen-ensemble}) is
generally an effective interaction, e.g., a screened Coulomb
interaction. Such an effective interaction was, for example, derived
in Refs.~\onlinecite{blanter97,aleiner02} within an interval of $\sim g$
levels around the Fermi energy of the dot. The authors used the
random--phase approximation (RPA) in the limit where the gas constant
$r_s \ll 1$. The effective interaction is then given by the unscreened
zero--momentum Fourier component of the Coulomb interaction plus a
short--range screened Coulomb interaction (whose exact form depends on
the dimensionality and geometry of the system). In general, for a
finite system this effective interaction also includes surface charge
terms.\cite{blanter97}  These can be eliminated by assuming, for
instance, periodic boundary conditions and will not be taken into
account in the sequel.

If the spin of the electrons is taken into account, the
Hamiltonian~(\ref{ensemble}) is replaced by
\begin{equation}
\label{ensemble-s}
H = \sum_{i,j;\sigma} h^{(0)}_{ij}a^\dagger_{i\sigma} a_{j\sigma}
 + {1\over 2} \sum_{ijkl  \atop \sigma\sigma'}
v_{ij;kl}a^\dagger_{i\sigma}  a^\dagger_{j\sigma'} a_{l\sigma'}
 a_{k\sigma} \;,
\end{equation}
where $\sigma = \pm$ describes a spin up/down electron and $|i\rangle$
denotes an orbital state. The one--body Hamiltonian $h^{(0)}$ is
spin--independent (i.e., there is no spin--orbit coupling).
The interaction matrix elements are also assumed to be
spin--independent as is the case for the Coulomb interaction.

In analogy to the transformation leading from Eq.~(\ref{ensemble}) to
Eq.~(\ref{eigen-ensemble}), the ensemble~(\ref{ensemble-s}) can be
written in terms of the orbital eigenfunctions $| \alpha \rangle$ of
the one--body Hamiltonian $h^{(0)}$. Since the transformation from the
basis of states $| i \rangle$ to that of the states $| \alpha \rangle$
does not involve the spin degree of freedom, the single--particle
energies $\epsilon_\alpha$ are spin--degenerate, and the interaction
matrix elements remain spin--independent. Thus, when the spin is
included, Eq.~(\ref{eigen-ensemble}) is replaced by
\begin{equation}
\label{eigen-ensemble-s}
H = \sum_{\alpha \sigma} \epsilon_{\alpha} a^\dagger_{\alpha \sigma}
a_{\alpha \sigma}+ {1 \over 2} \sum_{\alpha \beta \gamma \delta \atop
\sigma \sigma'} v_{\alpha \beta; \gamma \delta} a^\dagger_{\alpha
\sigma} a^\dagger_{\beta \sigma'} a_{\gamma \sigma'} a_{\delta \sigma}
\;.
\end{equation}
In contrast to the spinless case where $H$ could be expressed in terms
of the antisymmetrized matrix elements $v^A$,
Eq.~(\ref{eigen-ensemble-s}) contains the full non--antisymmetrized 
interaction matrix elements.

We now work out the first and second moments of $v_{\alpha \beta;
  \gamma \delta}$ by averaging over the eigenfunctions
$|\alpha\rangle$.

\subsection{Mean Values}

We first calculate the mean value $\bar v_{\alpha \beta; \gamma
  \delta}$ of the two--body matrix elements, invoking orthogonal or
unitary invariance. In the orthogonal case there are three invariants:
$\delta_{\alpha \gamma} \delta_{\beta \delta}$, $\delta_{\alpha
  \delta} \delta_{\beta \gamma}$, and $\delta_{\alpha \beta}
\delta_{\gamma \delta}$. In the unitary case, only the first two
invariants exist. We consider first the case of antisymmetrized matrix
elements and include spin below. For the antisymmetrized element
$v^A_{\alpha \beta; \gamma \delta} \equiv v_{\alpha \beta; \gamma
  \delta} - v_{\alpha \beta; \delta \gamma}$, the first two invariants
combine to $\delta_{\alpha \gamma} \delta_{\beta \delta} -
\delta_{\alpha \delta} \delta_{\beta \gamma}$, while the third one
gives no contribution. We conclude (for both the orthogonal and
unitary symmetries) that
\begin{equation}\label{average}
\bar v^A_{\alpha \beta; \gamma \delta} = v_0 (\delta_{\alpha \gamma}
\delta_{\beta \delta} - \delta_{\alpha \delta} \delta_{\beta \gamma})
\;.
\end{equation}
The average interaction in (\ref{eigen-ensemble}) is then given by
\begin{equation}\label{average-int}
\bar v = {1 \over 2} v_0 (\hat n^2 - \hat n) \;,
\end{equation}
where $\hat n$ is the particle--number operator. For a fixed number
$n$ of fermions, the average interaction is simply a constant $v_0
(n^2 - n)/2$. For a quantum dot, the zero--momentum Fourier component
of the Coulomb interaction is unscreened and its contribution to $v_0$
is the charging energy $e^2/C$ where $C$ is the dot's capacitance. The
contribution of the screened short--range Coulomb interaction can be
estimated using the diagrammatic expansion and is of order $\Delta$.
For finite $g$, there are corrections of order $1/g$.

For the case with spin, all three (two) orthogonal (unitary)
invariants now contribute to the value of the average interaction, 
\begin{equation}\label{average-s}
\bar v_{\alpha \beta; \gamma \delta} = v_0 \delta_{\alpha \gamma}
\delta_{\beta \delta} + J_s \delta_{\alpha \delta} \delta_{\beta \gamma}
+J_c \delta_{\alpha \delta} \delta_{\beta \gamma} \;.
\end{equation}
Here $v_0$, $J_s$ and $J_c$ are constants and the last term on the
r.h.s.~is absent for the unitary case. Using Eq.~(\ref{average-s}) in
Eq.~(\ref{eigen-ensemble-s}) we obtain an average interaction of the
form
\begin{equation}
\label{average-int-s}
\bar v = {1 \over 2}(v_0 - J_s/2) \hat n^2 - (v_0/2 -J_s) \hat n -
J_s{\bf \hat S}^2 + J_c \hat T^\dagger \hat T\;.
\end{equation}
In Eq.~(\ref{average-int-s}), ${\bf \hat S} = {1\over 2} \sum_\alpha
\sum_{\sigma \sigma'} a^\dagger_{\alpha \sigma}
{\hat {\bf \sigma}}_{\sigma \sigma'} a_{\alpha \sigma'}$ is the total
spin operator of the dot, where ${\hat {\bf \sigma}}$ is the vector of
the three $2\times 2$ Pauli matrices. The operator $\hat
T^\dagger=\sum_{\alpha} a^\dagger_{\alpha +} a^\dagger_{\alpha -}$
creates coherent pairs of spin up--down electrons. Adding the average
interaction (\ref{average-int-s}) to the one--body part of
Eq.~(\ref{eigen-ensemble-s}) and taking the limit $g \to \infty$,
we obtain the universal Hamiltonian of the quantum dot, with $v_0 =
\bar v_{\alpha \beta; \alpha \beta}$, $J_s = \bar v_{\alpha \beta;
  \beta \alpha}$, and $J_c = \bar v_{\alpha \alpha; \beta \beta}$ the
direct, exchange and the Cooper channel interaction strengths,
respectively.\cite{kurland00,aleiner02}  We note, however, that
Eq.~(\ref{average-int-s}) is valid beyond this limit. For finite
values of $g$, corrections to the various constants can be obtained by
the diagrammatic expansion in $1/g$. Unlike the charging energy term
(which for a fixed number of electrons is simply a constant), the
exchange and Cooper channel interactions are represented by
non--trivial operators.

\subsection{Second Moments}

The second moments of the interaction matrix elements depend upon the
symmetry of the ensemble, and upon the generic form of the two--body
interaction. It is easier to calculate first the second moments of the
non--antisymmetrized matrix elements $v_{\alpha \beta; \gamma \delta}$
for the two--body ensemble in the presence of spin degrees of freedom.
The spinless case is then obtained by considering the corresponding
moments of the antisymmetrized matrix elements $v^A_{\alpha \beta;
  \gamma \delta}$.

To simplify the notation, we assume in the following that the average
interaction has been separated out and put $\bar v_{\alpha \beta;
\gamma \delta} = 0$.

\subsubsection{Orthogonal induced two--body ensembles}
\label{ortho}

In the case of orthogonal symmetry the two--body interaction matrix
$v_{\alpha\beta;\gamma\delta}$ is real and symmetric, $v_{\alpha
  \beta; \gamma \delta} = v^*_{\alpha \beta; \gamma \delta} =
v_{\gamma \delta; \alpha \beta}$. Depending on the form of the
interaction, we distinguish three induced ensembles.

\noindent
(i) For a schematic effective interaction that is a $\delta$--function
(i.e., contact interaction), the matrix element $v_{\alpha \beta;
  \gamma \delta} = V \Delta \int d\bf r \psi_\alpha(\bf r) \psi_\beta
(\bf r) \psi_\gamma(\bf r) \psi_\delta(\bf r)$ ($V$ is the volume and
$\Delta$ is the single--particle mean--level spacing) is invariant
under all $24$ permutations of the four indices $\alpha \beta \gamma
\delta$. An ensemble of two--body matrix elements that satisfies the
orthogonal invariance condition and is consistent with the symmetries
of $v_{\alpha \beta; \gamma \delta}$, is expected to have to leading order in $1/g$ (see below) a second
moment of the form~\cite{mirlin}
\begin{equation}
\label{corr-delta}
\overline{ v_{\alpha \beta; \gamma \delta} v_{\mu \nu; \rho \sigma}} =
3u^2(\delta_{\alpha \mu} \delta_{\beta \nu} \delta_{\gamma \rho}
\delta_{\delta \sigma}  + \ldots) \;
\end{equation}
where $\ldots$ stands for the sum of terms obtained by all $23$
permutations of $\mu \nu \rho \sigma$.  Non--linear sigma model
calculations,\cite{mirlin00}  as well as diagrammatic calculations in
weakly disordered  systems using a screened Coulomb interaction, show
that $u^2 \propto \Delta^2/g^2$ (see below). We have written down only
invariants that arise by pairing indices appearing in different
matrix elements. Invariance requirements alone would also allow terms
involving Kronecker symbols like $\delta_{\alpha \beta}$ etc. However,
in diffusive systems such invariant terms are at least of third order
in $1/g$ and are, therefore, ignored. An analogous statement applies
to all the other induced two--body ensembles listed below, both for
the orthogonal and the unitary cases, and will not be repeated there.

For a diffusive dot, the quantity $u^2$ in Eq.~(\ref{corr-delta}) can
be calculated in the diagrammatic approach. If the energy difference
between any pair of levels from the set $\alpha \beta \gamma \delta$
is smaller than the Thouless energy $g \Delta$,
Eq.~(\ref{corr-delta}) is found with~\cite{blanter97,ag}
\begin{equation}
\label{variance}
u^2 =  \int d{\bf r} \int d{\bf r'} \ \Pi^2({\bf r},{\bf r'})/V^2 =
c {\Delta^2  \over g^2}\;,
\end{equation}
independently of the energy differences of the single--particle
levels. Here $\Pi( {\bf r}_1, {\bf r}_2)$ $=$ $(V\Delta/\pi) \sum_{\bf
  Q \neq 0} \phi_{\bf Q}({\bf r}_1) \phi_{\bf Q}({\bf r}_2) /D {\bf
  Q}^2$ is the diffusion propagator in the finite dot, with $\phi_{\bf
  Q}$ the eigenfunction of the diffusion operator corresponding to the
eigenvalue $D {\bf Q}^2$. The proportionality constant $c$ in
Eq.~(\ref{variance}) depends on the geometry of the dot. For a cubic
dot of length $L$ in $d$ dimensions, $c ={4 \over \pi^4} \sum_{{\bf n}
  \neq 0} {1 / |\bf n|^4}$ where $\bf n$ is a vector of $d$
non--negative integers (and the dimensionless conductance $g$ is
defined by $g \Delta = 2 \pi \hbar D/ L^2$).\cite{ag}  For a 2D
circular dot of radius $R$, $c= 2 [\sum_{l,m} x_{l,m}^{-4} ]^{1/2}
\approx 0.67$ where $x_{l,m}$ are the zeros of the derivative of
the Bessel function of order $l$ (and the dimensionless conductance
is defined by $g \Delta = 2 \pi \hbar D/R^2$).\cite{alhassid02a}
In the general case where some of the energy differences between pairs
of levels are greater than $g\Delta$, the coefficient $c$ in
Eq.~(\ref{variance}) is no longer constant but depends on these energy
differences. 

Eq.~(\ref{variance}) can be extended to a ballistic dot. This is done
with the help of a supersymmetric sigma model, obtained through the
addition of weak disorder with finite correlation
length.\cite{blanter01} In the ballistic case~\cite{alhassid02}
\begin{equation}
u^2= c^\prime {\Delta^2 \over g^2} \ln(c'' g)
\;,
\end{equation} 
where $g=\pi (n/2)^{1/2}$ is the ballistic Thouless conductance, $n$
is the number of electrons, and $c^\prime = 3/4$ is a
geometry--independent constant. The constant $c''$ depends on the
geometry of the dot. For a circular ballistic dot, $c'' \approx 0.81$.

From Eq.~(\ref{corr-delta}), the variances of off--diagonal and
diagonal matrix elements are given by ($\alpha,\beta,\gamma,\delta$
are all different)
\begin{eqnarray}
\sigma^2(v_{\alpha \beta; \gamma \delta}) =  3 u^2 &\;,&\;\;
\sigma^2(v_{\alpha \beta; \alpha \beta})  = 12 u^2 \;,\;\; \nonumber \\
\sigma^2(v_{\alpha \alpha; \alpha \alpha}) & = & 72 u^2\;.
\end{eqnarray}

For spinless fermions, we have $v^A_{\alpha\beta; \gamma \delta} =
  v_{\alpha \beta; \gamma \delta}-v_{\alpha \beta; \delta\gamma} = 0$
because of the symmetries satisfied by the $v$'s. Therefore, there is
no non--trivial ensemble of antisymmetrized matrix elements that
corresponds to Eq.~(\ref{corr-delta}), and the
ensemble~(\ref{corr-delta}) exists only when the spin degrees of
freedom are taken into account. 

\begin{widetext}
\noindent (ii) For a generic two--body local symmetric interaction
$v({\bf  r_1},{\bf r_2})= v({\bf r_2},{\bf r_2})$, the matrix elements
satisfy the relations
\begin{equation}\label{v-local}
v_{\alpha\beta;\gamma\delta} = v_{\beta\alpha;\delta\gamma}=v_{\gamma\delta;\alpha\beta}=v_{\delta\gamma;\beta\alpha}=
v_{\gamma\beta;\alpha\delta}=v_{\beta\gamma;\delta\alpha}=v_{\alpha\delta;\gamma\beta}=v_{\delta\alpha;\beta\gamma}\;.
\end{equation}
The second moment of the ensemble is expected to have the form
\begin{eqnarray}
\label{corr-local}
\overline{ v_{\alpha \beta; \gamma \delta} v_{\mu \nu; \rho \sigma}} =
u^2(\delta_{\alpha \mu} \delta_{\beta \nu} \delta_{\gamma \rho}
\delta_{\delta \sigma }  +
\delta_{\alpha \nu} \delta_{\beta \mu} \delta_{\gamma \sigma}
\delta_{\delta \rho} +
\delta_{\alpha \rho} \delta_{\beta \sigma} \delta_{\gamma \mu}
\delta_{\delta \nu} +
\delta_{\alpha \sigma }\delta_{\beta \rho} \delta_{\gamma \nu}
\delta_{\delta \mu} + \nonumber \\
\delta_{\alpha \rho} \delta_{\beta nu} \delta_{\gamma \mu}
\delta_{\delta \sigma} +
\delta_{\alpha \nu} \delta_{\beta \rho} \delta_{\gamma \sigma}
\delta_{\delta mu} +
\delta_{\alpha \mu} \delta_{\beta \sigma} \delta_{\gamma \rho}
\delta_{\delta \nu} +
\delta_{\alpha \sigma} \delta_{\beta \mu} \delta_{\gamma \nu}
\delta_{\delta \rho}) \;,
\end{eqnarray}
%\end{widetext}
where the eight terms correspond to the eight permutations of $\mu \nu
\rho \sigma$ that leave $v_{\mu \nu \rho \sigma}$ invariant (see
Eqs.~(\ref{v-local})). The form~(\ref{corr-local}) is also obtained in
the diagrammatic approach to order $1/g^2$ when an RPA two--body
screened interaction $v(\bf r_1,\bf r_2)$ is used. In the limit $r_s
\ll 1$, some of the diagrams that do contribute for a
$\delta$--function interaction, give a negligible contribution for a
local finite--range interaction,\cite{mirlin} this leading to
Eq.~(\ref{corr-local}) instead of Eq.~(\ref{corr-delta}). The factor
$u^2$ is still given approximately by Eq.~(\ref{variance}), but now
\begin{eqnarray}
\sigma^2(v_{\alpha\beta;\gamma\delta})  =   u^2 \;,\;\;
\sigma^2(v_{\alpha\beta;\alpha\beta})  = 4 u^2 \;,\;\; 
\sigma^2(v_{\alpha\beta;\beta\alpha}) =  2 u^2 \;,\;\;
\sigma^2(v_{\alpha\alpha;\alpha\alpha})  = 8 u^2 \ .
\label{moments}
\end{eqnarray}
The direct ($v_{\alpha \beta; \gamma \delta}$) and exchange
($v_{\alpha \beta; \delta \gamma}$) matrix elements are uncorrelated.
The ensemble of antisymmetrized matrix elements $v^A_{\alpha \beta;
  \gamma \delta}$ follows from Eq.~(\ref{corr-local}) using
$\overline{v^A_{\alpha \beta; \gamma \delta} v^A_{\mu \nu; \rho
    \sigma}} = \overline{v_{\alpha \beta; \gamma \delta} v_{\mu \nu;
    \rho \sigma}} + \overline{v_{\alpha \beta; \delta \gamma} v_{\mu
    \nu; \sigma \rho}} - \overline{v_{\alpha \beta; \gamma \delta}
  v_{\mu \nu; \sigma \rho}} - \overline{v_{\alpha \beta; \delta
    \gamma} v_{\mu \nu; \rho \sigma}}$.
We find
%\begin{widetext}
\begin{eqnarray}
\label{corr-local-a}
\overline{ v^A_{\alpha \beta; \gamma \delta} v^A_{\mu \nu; \rho
    \sigma}} = 2u^2(\delta_{\alpha \mu} \delta_{\beta \nu}
    \delta_{\gamma \rho} \delta_{\delta \sigma} - \delta_{\alpha \mu}
    \delta_{\beta \nu} \delta_{\gamma \sigma} \delta_{\delta \rho} +
    \delta_{\alpha \nu} \delta_{\beta \mu} \delta_{\gamma \sigma}
    \delta_{\delta \rho} - \delta_{\alpha \nu} \delta_{\beta \mu}
    \delta_{\gamma \rho} \delta_{\delta \sigma} \nonumber \\
+ \delta_{\alpha \rho} \delta_{\beta \sigma} \delta_{\gamma \mu}
    \delta_{\delta \nu} - \delta_{\alpha \sigma} \delta_{\beta \rho}
    \delta_{\gamma \mu} \delta_{\delta \nu} + \delta_{\alpha \sigma}
    \delta_{\beta \rho} \delta_{\gamma \nu} \delta_{\delta \mu} -
    \delta_{\alpha \rho} \delta_{\beta \sigma} \delta_{\gamma \nu}
    \delta_{\delta \mu} ) \nonumber \\
+ u^2 (\delta_{\alpha \rho} \delta_{\beta \nu} \delta_{\gamma \mu}
    \delta_{\delta \sigma} - \delta_{\alpha \sigma} \delta_{\beta \nu}
    \delta_{\gamma \mu} \delta_{\delta \rho} + \delta_{\alpha \nu}
    \delta_{\beta \rho} \delta_{\gamma \sigma} \delta_{\delta \mu} -
    \delta_{\alpha \nu} \delta_{\beta \sigma} \delta_{\gamma \rho}
    \delta_{\delta \mu} \nonumber \\
+ \delta_{\alpha \mu} \delta_{\beta \sigma} \delta_{\gamma \rho}
    \delta_{\delta \nu} - \delta_{\alpha \mu} \delta_{\beta \rho}
    \delta_{\gamma \sigma} \delta_{\delta \nu} + \delta_{\alpha
    \sigma} \delta_{\beta \mu} \delta_{\gamma \nu} \delta_{\delta
    \rho} - \delta_{\alpha \rho} \delta_{\beta \mu} \delta_{\gamma
    \nu} \delta_{\delta \sigma} \nonumber \\
+ \delta_{\alpha \sigma} \delta_{\beta \nu} \delta_{\gamma \rho}
    \delta_{\delta \mu} - \delta_{\alpha \rho} \delta_{\beta \nu}
    \delta_{\gamma \sigma} \delta_{\delta \mu} + \delta_{\alpha \nu}
    \delta_{\beta \sigma} \delta_{\gamma \mu} \delta_{\delta \rho} -
    \delta_{\alpha \nu} \delta_{\beta \rho} \delta_{\gamma \mu}
    \delta_{\delta \sigma} \nonumber \\ + \delta_{\alpha \mu}
    \delta_{\beta \rho} \delta_{\gamma \nu} \delta_{\delta \sigma} -
    \delta_{\alpha \mu} \delta_{\beta \sigma} \delta_{\gamma \nu}
    \delta_{\delta \rho} + \delta_{\alpha \rho} \delta_{\beta \mu}
    \delta_{\gamma \sigma} \delta_{\delta \nu} - \delta_{\alpha
    \sigma} \delta_{\beta \mu} \delta_{\gamma \rho} \delta_{\delta
    \nu}) \;.
\end{eqnarray}
%\end{widetext}
In particular
\begin{eqnarray}
\label{variances-local}
\sigma^2(v^A_{\alpha\beta;\gamma\delta}) =  2 u^2
\;,\;\;\sigma^2(v^A_{\alpha\beta;\alpha\beta})  = 6 u^2 \;.
\end{eqnarray}

\noindent (iii) The matrix elements of a generic non--local two--body
interaction  $v({\bf r_1}, {\bf r_2};{\bf r'_1}, {\bf r'_2})= v({\bf
  r_2}, {\bf  r_1};{\bf r'_2}, {\bf r'_1})$ (we define $v({\bf r_1},
{\bf r_2};{\bf  r'_1}, {\bf r'_2})\equiv \langle {\bf r_1}, {\bf r_2}
| v | {\bf r'_1},  {\bf r'_2}\rangle$) satisfy
\begin{equation}
\label{v-nonlocal}
v_{\alpha\beta;\gamma\delta} = v_{\beta\alpha;\delta\gamma}=v_{\gamma\delta;\alpha\beta}=v_{\delta\gamma;\beta\alpha}\;.
\end{equation}
Therefore, we expect for the ensemble average
\begin{eqnarray}
\label{corr-goe}
\overline{ v_{\alpha \beta; \gamma \delta} v_{\mu \nu; \rho \sigma}} =
u^2(\delta_{\alpha \mu } \delta_{\beta \nu} \delta_{\gamma \rho}
\delta_{\delta \sigma} +
\delta_{\alpha \nu} \delta_{\beta \mu} \delta_{\gamma \sigma}
\delta_{\delta \rho} +
\delta_{\alpha \rho} \delta_{\beta \sigma} \delta_{\gamma \mu}
\delta_{\delta \nu} +
\delta_{\alpha \sigma} \delta_{\beta \rho} \delta_{\gamma \nu}
\delta_{\delta \mu}) \;,
\end{eqnarray}
i.e., the two--body Gaussian orthogonal ensemble. The corresponding
variances are given by
\begin{eqnarray}
\sigma^2(v_{\alpha\beta;\gamma\delta}) =  u^2 \;,\;\;
\sigma^2(v_{\alpha\beta;\alpha\beta})  = 2 u^2 \;,\;\;
\sigma^2(v_{\alpha\beta;\beta\alpha})  = 2 u^2 \;,\;\;
\sigma^2(v_{\alpha\alpha;\alpha\alpha})  = 4 u^2 \;.
\end{eqnarray}

For the antisymmetrized interaction matrix elements we expect
accordingly
\begin{eqnarray}
\label{corr-goe-a}
\overline{ v^A_{\alpha \beta; \gamma \delta} v^A_{\mu \nu; \rho
    \sigma}} = 2u^2(\delta_{\alpha \mu} \delta_{\beta \nu}
    \delta_{\gamma \rho} \delta_{\delta \sigma} - \delta_{\alpha \mu}
    \delta_{\beta \nu} \delta_{\gamma \sigma} \delta_{\delta \rho} +
    \delta_{\alpha \nu} \delta_{\beta \mu} \delta_{\gamma \sigma}
    \delta_{\delta \rho} - \delta_{\alpha \nu} \delta_{\beta \mu}
    \delta_{\gamma \rho} \delta_{\delta \sigma} \nonumber \\
+ \delta_{\alpha \rho} \delta_{\beta \sigma} \delta_{\gamma \mu}
    \delta_{\delta \nu} - \delta_{\alpha \sigma} \delta_{\beta \rho}
    \delta_{\gamma \mu} \delta_{\delta \nu} + \delta_{\alpha \sigma}
    \delta_{\beta \rho} \delta_{\gamma \nu} \delta_{\delta \mu} -
    \delta_{\alpha \rho} \delta_{\beta \sigma} \delta_{\gamma \nu}
    \delta_{\delta \mu}) \;.
\end{eqnarray}
%\end{widetext}

In particular
\begin{eqnarray}\label{variances-goe}
\sigma^2(v^A_{\alpha\beta;\gamma\delta}) =  2 u^2
\;,\;\;\sigma^2(v^A_{\alpha\beta;\alpha\beta})  = 4 u^2 \;.
\end{eqnarray}
These relations should be compared with Eqs.~(\ref{variances-local}).
It is not clear whether there is an interaction model for which
disorder averaging leads to the ensemble (\ref{corr-goe-a}), although
this ensemble seems to be generic for non--local interactions.

\subsubsection{Unitary induced two--body ensembles}\label{unitary}

For the unitary symmetry the two--body interaction matrix is complex
Hermitean, $v_{\alpha\beta;\gamma\delta} =
 v^*_{\gamma\delta;\alpha\beta}$. There are two possible two--body
ensembles:

%\begin{widetext}
\noindent (i) The matrix elements $v_{\alpha\beta;\gamma\delta}$
 satisfy the relations
\begin{equation}
\label{v-delta-u}
v_{\alpha\beta;\gamma\delta} = v_{\beta\alpha;\gamma\delta}= v_{\alpha\beta;\delta\gamma} = v_{\beta\alpha;\delta\gamma}=
v^*_{\gamma\delta;\alpha\beta} = v^*_{\gamma\delta;\beta\alpha}= v^*_{\delta\gamma;\alpha\beta} = v^*_{\gamma\delta;\beta\alpha} \;.
\end{equation}
The  ensemble consistent with relations~(\ref{v-delta-u}) is
\begin{equation}
\label{corr-delta-u}
\overline{ v^*_{\alpha \beta; \gamma \delta} v_{\mu \nu; \rho \sigma}}
= 2u^2(\delta_{\alpha \mu} \delta_{\beta \nu} \delta_{\gamma \rho}
\delta_{\delta \sigma} +
\delta_{\alpha \nu} \delta_{\beta \mu} \delta_{\gamma \rho}
\delta_{\delta \sigma} +
\delta_{\alpha \mu} \delta_{\beta \nu} \delta_{\gamma \sigma}
\delta_{\delta \rho} +
\delta_{\alpha \nu} \delta_{\beta \mu} \delta_{\gamma \sigma}
\delta_{\delta \rho}) \;.
\end{equation}
In the disorder basis such an ensemble is realized for a
$\delta$--function interaction with $u^2$ given by
Eq.~(\ref{variance}). Variances of diagonal and off--diagonal elements
are given by
\begin{eqnarray}
\sigma^2(v_{\alpha \beta; \gamma \delta}) = 2 u^2 \;,\;\;
\sigma^2(v_{\alpha \beta; \alpha \beta})  = 2 u^2 \;,\;\;
\sigma^2(v_{\alpha \beta; \beta \alpha})  = 2 u^2 \;,\;\;
\sigma^2(v_{\alpha \alpha; \alpha \alpha})  = 8 u^2\;.
\end{eqnarray}
Relations~(\ref{v-delta-u}) lead to $v^A_{\alpha \beta; \gamma \delta}
= 0$, so there is no non--trivial ensemble of antisymmetrized matrix
elements that corresponds to Eq.~(\ref{corr-delta-u}).

\noindent (ii) The matrix elements satisfy relations that are typical
for a symmetric two--body interaction (local or non--local)
\begin{equation}
\label{v-local-u}
v_{\alpha\beta;\gamma\delta} = v_{\beta\alpha;\delta\gamma}= v^*_{\gamma\delta;\alpha\beta} = v^*_{\delta\gamma;\beta\alpha}\;.
\end{equation}
A corresponding induced two--body random matrix ensemble obeys
\begin{eqnarray}
\label{corr-gue}
\overline{ v^*_{\alpha \beta; \gamma \delta} v_{\mu \nu; \rho \sigma}}
= u^2 (\delta_{\alpha \mu} \delta_{\beta \nu} \delta_{\gamma \rho}
\delta_{\delta \sigma} +
\delta_{\alpha \nu} \delta_{\beta \mu} \delta_{\gamma \sigma}
\delta_{\delta \rho}) \;,
\end{eqnarray}
i.e., the two--body Gaussian unitary ensemble. Such an ensemble is
realized in the disorder basis by an RPA two--body screened
interaction in the limit $r_s \ll 1$, with $u^2$ given approximately
by Eq.~(\ref{variance}). The variances of the matrix elements are
\begin{eqnarray}
\sigma^2(v_{\alpha \beta; \gamma \delta}) =  u^2 \;,\;\;
\sigma^2(v_{\alpha \beta; \alpha \beta}) = u^2 \;\;\;
\sigma^2(v_{\alpha \beta;\beta\alpha }) = u^2\;,\;\;
\sigma^2(v_{\alpha \alpha; \alpha \alpha}) = 2 u^2\;.
\end{eqnarray}
For the antisymmetrized interaction we find
\begin{eqnarray}
\label{corr-gue-a}
\overline{ v^A_{\alpha \beta; \gamma \delta}v^A_{\mu \nu; \rho
    \sigma}} = 2u^2 (\delta_{\alpha \mu} \delta_{\beta \nu}
    \delta_{\gamma \rho} \delta_{\delta \sigma} - \delta_{\alpha \mu}
    \delta_{\beta \nu} \delta_{\gamma \sigma} \delta_{\delta \rho} +
    \delta_{\alpha \nu} \delta_{\beta \mu} \delta_{\gamma \sigma}
    \delta_{\delta \rho} - \delta_{\alpha \nu} \delta_{\beta \mu}
    \delta_{\gamma \rho} \delta_{\delta \sigma}) \;,
\end{eqnarray}
and in particular
\begin{eqnarray}
\sigma^2(v^A_{\alpha \beta; \gamma \delta}) = 2 u^2
\;,\;\;\sigma^2(v^A_{\alpha \beta; \alpha \beta})  = 2 u^2 \;.
\end{eqnarray}
\end{widetext}

In the orthogonal case there was a third ensemble, but in the unitary
case this third ensemble coincides with the second one because the
unitary symmetry relations~(\ref{v-local-u}) are common to both local
and non--local interactions.

We reiterate that in the spinless case, the only non--trivial induced
ensembles are (ii) and (iii) in the orthogonal case
(Section~\ref{ortho}), and (ii) in the unitary case
(Section~\ref{unitary}).

\subsection{Induced versus true two--body ensembles}
\label{ind}

To simplify the discussion, we disregard the average part of the
interaction in the ensembles~(\ref{eigen-ensemble}) and
(\ref{eigen-ensemble-s}) and put $\bar v_{\alpha \beta; \gamma \delta}
= 0$.

So far, we have calculated only the second moments of the two--body
interaction. What about higher moments? Inspection of the relevant
diagrams shows that all higher--order cumulants contribute terms of
order $1/g^k$ with $k \geq 3$ and are, thus, negligible in the present
context. This does not necessarily imply that the induced two-body
random ensembles are Gaussian. (It is conceivable that the $k$th
cumulants are systematically of order $1/g^k$. Rescaling the matrix
elements by the factor $g$ would then yield a variable which is not
Gaussian). On the other hand, numerical results in weakly
diffusive systems do suggest that the interaction matrix elements have Gaussian distributions.\cite{cohen99}  This can be attributed
 to a central--limit theorem: each matrix element is an integral
over a product of four wave functions, and each wave function is a
continuous random variable.

  Our results might suggest that instead of the induced random two--body
ensembles studied so far, we may use true random two--body ensembles.
In the latter, the matrix elements $v_{ij;kl}$ are
Gaussian--distributed random variables with mean value zero and second
moments given by Eqs.~(\ref{corr-delta}), (\ref{corr-local}),
(\ref{corr-goe}), (\ref{corr-delta-u}) or (\ref{corr-gue}), as the case
may be. These relations apply in any single--particle basis. In
contrast to the induced random two--body ensembles, the statistical
properties of the true two--body random ensembles are inherent rather
than induced by the randomness of the one--body part of $H$. Are these
two sets of ensembles pairwise equivalent? Are we allowed to replace a
Hamiltonian with a random one--body and a fixed two--body interaction
by a generic Hamiltonian containing both a random one--body and the
random two--body interaction just defined?  The answer is no, even
though both Hamiltonians yield the same distributions for the
two--body matrix elements. This is because in the induced two--body
ensembles, randomness is due to the eigenfunctions $|\alpha\rangle$ of
$h^{(0)}$. Therefore, the two--body matrix elements $v_{\alpha \beta;
  \gamma \delta}$ are correlated with all other expressions that
depend on the $|\alpha\rangle$'s while the matrix elements of the true
two--body random ensemble are independent random variables that are
not correlated with other parts of the Hamiltonian. Whenever an
observable contains in addition to the $v_{\alpha \beta; \gamma
  \delta}$'s other expressions which depend upon the
$|\alpha\rangle$'s, the ensemble averages of this observable taken
over the induced two--body ensemble and over the fully random
two--body ensemble, will differ. However, in the numerical studies of
the statistics of HF matrix elements reported below, such correlated
terms do not occur, and we use the fully random two--body ensemble.

\section{The Hartree-Fock approach to disordered systems: fixed number
  of electrons}

The theoretical treatment of a disordered system with interactions
poses severe difficulties. This is why we use the HF approximation. We
express observables in terms of the HF single--particle eigenvalues
and eigenfunctions. We determine the latter by using the HF
approximation for every realization of disorder, i.e., for every
realization of a single--particle random--matrix spectrum
$\epsilon_\alpha$ and random interaction matrix elements $v_{\alpha
\beta; \gamma \delta}$ in Eq.~(\ref{eigen-ensemble}). This generates
an ensemble of HF eigenvalues and HF eigenfunctions. We use the
ensemble to calculate average values of observables as well as
fluctuation properties (e.g., variances and/or distributions). In
particular, we are interested in the statistical properties of the
interaction matrix elements in the HF basis, and of the
single--particle HF energies and wave functions.

All our calculations are done for spinless electrons. Therefore, we
use the antisymmetrized matrix elements defined in the previous
Section. We disregard the average matrix elements which (for a fixed
number of electrons) would add only a constant to the Hamiltonian. 
To simplify the notation we omit the superscript $A$. It goes without
saying that inclusion of spin degrees of freedom would be highly
desirable but would obviously complicate the calculations considerably.
The present paper is, thus, an exploratory study into the use of the
HF approximation for a disordered system with interactions. Similar
studies were previously carried out in the context of specific models,
using the Coulomb interaction or a Hubbard--like short--range
interaction,\cite{levit99,walker99,cohen99}  while our work is based 
on the generic many--body random ensembles introduced in
Section~\ref{ran}.

In Section~\ref{HF}, we discuss the HF approximation in the context of
the induced many--body ensembles. Sections~\ref{mom} and
\ref{HF-elements} are devoted to the study of the statistical
properties of the interaction matrix elements in the self--consistent
HF basis. In particular, in Section~\ref{mom} we derive perturbative
analytic expressions for the first two moments of the HF interaction
matrix elements (assuming small $u$ and a fixed single--particle
spectrum), while in Section~\ref{HF-elements} we study numerically the
statistical properties of the HF interaction matrix elements. Finally,
in Sections~\ref{HF-spect} and \ref{HF-wavefunctions} we study the
statistical properties of the HF single--particle spectrum and wave
functions.

\subsection{Hartree-Fock approximation for an interacting disordered
system}\label{HF}

In the eigenbasis of $h^{(0)}$, the single--particle HF Hamiltonian
for $n$ electrons is given by
\begin{equation}\label{HF-Ham}
h_{\alpha \gamma} = \epsilon_\alpha \delta_{\alpha \gamma} +
\sum_{\beta \delta} v_{\alpha \beta; \gamma \delta} \rho_{\delta
  \beta} \;,
\end{equation}
where $\rho$ is the density matrix 
\begin{equation}\label{density-matrix}
\rho_{\delta \beta} = \sum_{l=1}^{n} \psi_l^*(\delta) \psi_l(\beta) \;.
\end{equation}
Here ${\bf \psi}_l = \sum_\alpha \psi_l(\alpha) |\alpha\rangle$ are
the lowest $n$ HF eigenstates. The latter satisfy
\begin{equation}\label{HF-equations}
\sum_\gamma h_{\alpha \gamma} \psi_l(\gamma) = \epsilon^{(n)}_l
\psi_l(\alpha)\;\;\;\; ({\rm for\; each}\; \alpha)
\end{equation}
where $\epsilon^{(n)}_l$ are the single--particle HF energies for $n$
electrons (for $n=1$ there are no interactions and $\epsilon^{(1)}_l =
\epsilon_l$). We note that the single--particle HF Hamiltonian
$h_{\alpha \gamma}$ depends on the number of electrons $n$, but for
simplicity of notation we do not indicate this dependence explicitly.

The HF equations~(\ref{HF-equations}) are self--consistent since the 
HF Hamiltonian itself is determined in terms of its $n$ lowest
eigenstates $\psi_l$ [see Eqs.~(\ref{HF-Ham}) and
(\ref{density-matrix})]. The HF equations are solved by iteration.
Starting from a particular realization of $\epsilon_\alpha$ and
$v_{\alpha\beta;\gamma\delta}$, we calculate the zeroth 
approximation $h_0$ to the HF Hamiltonian, using Eq.~(\ref{HF-Ham})
and the density matrix $\rho^{(0)}$ given by Eq.~(\ref{density-matrix}),
with $\psi^{(0)}_l (\beta) = \delta_{l \beta}$ (i.e., the eigenstates
of $h^{(0)}$). Diagonalization of $h_0$ yields a new set of
single--particle eigenvalues $\epsilon_{1; \sigma}$ and wave functions
$\psi^{(1)}_\sigma(\alpha)$. The latter are used to construct the
density matrix $\rho^{(1)}$ from Eq.~(\ref{density-matrix}). All this
yields the one--body Hamiltonian
\begin{equation}
h_{1; \sigma \tau} = \epsilon_{1; \sigma} \delta_{\sigma \tau} +
\sum_{\mu \nu} v_{\sigma \mu; \tau \nu} \rho^{(1)}_{\nu \mu} \;.
\end{equation}
Diagonalizing $h_1$ we find its eigenvalues $\epsilon_{2; \sigma}$
and eigenstates $\psi^{(2)}_\sigma$. The latter are used in turn to
construct a density matrix $\rho^{(2)}$, etc. In general, $\rho^{(2)}
\neq \rho^{(1)}$ and likewise for the eigenvalues, and the iteration
continues until the procedure converges to a self--consistent solution
for $\psi_l$ and $\epsilon^{(n)}_l$. To ensure convergence, it is
often necessary to replace, e.g., $\rho^{(2)}$ by $\tilde \rho^{(2)} =
(1-\lambda) \rho^{(1)} + \lambda \rho^{(2)}$, where $\lambda$ ($0 <
\lambda \leq 1$) is a suitably chosen parameter.

We define the HF basis as that basis in which the HF Hamiltonian is
diagonal, $h_{\sigma \tau}= \epsilon^{(n)}_\sigma \delta_{\sigma
  \tau}$ and $\rho_{\sigma \tau} = \delta_{\sigma \tau}
\Theta(n-\sigma)$ where $\Theta(x) = 1$ for $x \geq 0$ 
and $\Theta(x) = 0$ for $x<0$. The HF eigenstates $\sigma$ are
arranged in ascending order of the HF single--particle energies
$\epsilon^{(n)}_\sigma$. The latter can be written as
\begin{equation}\label{HF-level}
\epsilon^{(n)}_\sigma = h^{(0)}_{\sigma \sigma} + \sum_{\tau = 1}^n
 v_{\sigma \tau; \sigma \tau}\;,
\end{equation}
where $h^{(0)}_{\sigma \sigma}=\sum_{\alpha}\epsilon_\alpha
 |\psi_\sigma(\alpha)|^2$ is a diagonal matrix element of the
single--particle Hamiltonian $h^{(0)}$ in the HF basis $\sigma$. We
emphasize that in general, $h^{(0)}_{\sigma \sigma} \neq
\epsilon_\sigma$.

\subsection{Moments of the Hartree--Fock matrix elements: perturbative
approach}\label{mom}

In Section~\ref{HF-elements} below, we report on numerical HF
calculations. The calculations are done for each realization of the
many--particle ensemble~(\ref{eigen-ensemble}), and statistics are
collected from different realizations. This is done for either a fixed
single--particle spectrum (where only the single--particle
eigenfunctions are varied) or for a random single--particle spectrum
(where both the single--particle energies and wave functions are
varied). In either case, the resulting distribution of the two--body
matrix elements in the HF basis turns out to be Gaussian. We compare
the numerical results for $n$ fermions with theoretical predictions
based upon perturbation theory. To this end, we compute in the present
Section the first and second moments of the two--body HF matrix
elements perturbatively in leading order in $u^2$ for a {\em fixed}
single--particle spectrum. We do so by using the statistics of the
true (rather than of the induced) two--body ensembles introduced in
Section~\ref{ran}. This is legitimate because we average expressions
that contain only the two--body matrix elements (and not other terms
depending upon the eigenfunctions $|\alpha \rangle$ of $h^{(0)}$, see
Section~\ref{ind}). We confine ourselves to the case of orthogonal
symmetry. We are particularly interested in the diagonal matrix
elements.

\subsubsection{Perturbation expansion}

We recall Eqs.~(\ref{HF-Ham}), (\ref{density-matrix}) and
(\ref{HF-equations}). To first order in $v$, the HF Hamiltonian is
obtained by substituting in Eq.~(\ref{HF-Ham}) the zeroth order
expression for the one--body density matrix given by
$\rho^{(0)}_{\delta \beta} = \delta_{\delta \beta} \Theta(n-\beta)$.
(We assume that the eigenvalues $\epsilon_\alpha$ are arranged in
ascending order). Thus,
\begin{equation}
\label{HF-Hamiltonian-1}
h_{0; \alpha \gamma} = \epsilon_\alpha \delta_{\alpha \gamma} +
\sum\limits_{\delta=1}^{n} v^{(0)}_{\alpha \delta; \gamma \delta}\;,
\end{equation}
where $v^{(0)}$ denotes a matrix element in the basis of eigenstates
$|\alpha \rangle$ of $h^{(0)}$.

Regarding $\sum\limits_{\delta=1}^{n} v^{(0)}_{\alpha \delta; \gamma
\delta}$ as a perturbation to the Hamiltonian $h^{(0)}$, we see that
a HF eigenstate is (to first order in $v^{(0)}$) given by
\begin{equation} \label{HF-wfunct}
|\alpha\rangle_{\rm HF} = |\alpha\rangle + \sum_{\mu \neq \alpha} 
r^{(1)}_{\alpha  \mu} |\mu\rangle \;,
\end{equation}
where 
\begin{equation}
\label{r1}
r^{(1)}_{\alpha \mu} = {1 \over \epsilon_\alpha - \epsilon_\mu}
\sum_{\delta} \Theta(n-\delta) v^{(0)}_{\mu \delta; \alpha \delta} \ .
\end{equation}
Using Eq.~(\ref{HF-wfunct}), we find that to second order in $v^{(0)}$
the matrix element $v_{\alpha \delta; \gamma \delta}$ in the HF basis
is given by 
\begin{widetext}
\begin{equation}\label{HF-element-g}
v_{\alpha\beta;\gamma\delta} = v^{(0)}_{\alpha\beta;\gamma\delta} +
\sum_{\mu \neq  \alpha} r^{(1)}_{\alpha \mu}v^{(0)}_{\mu
  \beta;\gamma\delta} + \sum_{\mu \neq  \beta} r^{(1)}_{\beta
  \mu}v^{(0)}_{\alpha\mu;\gamma\delta} +  \sum_{\mu \neq  \gamma}
r^{(1)}_{\gamma \mu}v^{(0)}_{\alpha \beta;\mu\delta} + \sum_{\mu \neq
  \delta} r^{(1)}_{\delta \mu}v^{(0)}_{\alpha\beta;\gamma\mu}\;.
\end{equation}

To calculate the variance of a HF matrix element $v_{\alpha \beta;
  \gamma \delta}$ to fourth order in $u$, we need to expand $v_{\alpha
  \beta; \gamma \delta}$ up to third order in $v^{(0)}$. This requires
an expansion of the HF state to second order in $v^{(0)}$,
\begin{equation}
|\alpha\rangle_{\rm HF} = \left( 1 - {1 \over 2} \sum_{\mu \neq
  \alpha} ({r_{\alpha \mu}^{(1)}})^2 \right) |\alpha\rangle +
  \sum_{\mu \neq \alpha} r^{(1)}_{\alpha  \mu} |\mu\rangle + \sum_{\mu
  \neq \alpha} r^{(2)}_{\alpha \mu}  |\mu\rangle \;,
\end{equation}
where $r^{(1)}$ and $r^{(2)}$ are of first and second order in
$v^{(0)}$, respectively. The matrix element is then expanded as
\begin{eqnarray}
v_{\alpha \beta; \gamma \delta} & = &\left(1 -{1 \over 2} \sum_{\mu
  \neq \alpha} ({r_{\alpha\mu}^{(1)}})^2 - {1 \over 2} \sum_{\mu \neq
  \beta} ({r_{\beta\mu}^{(1)}})^2 - {1 \over 2} \sum_{\mu \neq \gamma}
  ({r_{\gamma \mu}^{(1)}})^2 -{1 \over 2} \sum_{\mu \neq \delta}
  ({r_{\delta\mu}^{(1)}})^2 \right) v^{(0)}_{\alpha \beta; \gamma
  \delta} \nonumber \\
&& + \left[ \sum_{\mu \neq \alpha} (r^{(1)}_{\alpha \mu} +
  r^{(2)}_{\alpha \mu}) v^{(0)}_{\mu \beta; \gamma \delta} + \ldots
  \right] +\left[ \sum_{\mu \neq \alpha; \nu \neq \beta}
  r^{(1)}_{\alpha \mu} r^{(1)}_{\beta \nu} v^{(0)}_{\mu \nu; \gamma
  \delta} + \ldots \right] \;,
\end{eqnarray}
where $r^{(1)}_{\alpha \mu}$ is given by Eq.~(\ref{r1}). To find
$r^{(2)}_{\alpha \mu}$ we have to expand the HF Hamiltonian to second
order in $v^{(0)}$ and use the perturbation expansion for the
single--particle wave functions up to second order. The HF Hamiltonian
$h^{(2)}$ is found to second order in $v^{(0)}$ by expanding the
one--body density to first order in $v^{(0)}$,
\begin{equation}
\rho_{\delta \beta} = \sum \limits_{l=l}^n \psi_l -\delta
\psi_l(\beta) = \delta_{\delta \beta} \Theta(n-\beta) +
[r^{(1)}_{\delta \beta} \Theta(n-\delta) + r^{(1)}_{\beta \delta}
\Theta(n-\beta)] (1 - \delta_{\beta \delta}) \;. 
\end{equation}
Substituting this into Eq.~(\ref{HF-Ham}) we find
\begin{eqnarray}
\label{HF-Hamiltonian-2}
h^{(2)}_{\alpha \gamma} & = & \epsilon_\alpha \delta_{\alpha \gamma} +
\sum_{\delta} v^{(0)}_{\alpha \delta; \gamma \delta} \Theta(n-\delta)
+ \sum_{\beta \delta} v^{(0)}_{\alpha \beta; \gamma \delta}
[r^{(1)}_{\delta \beta} \Theta(n-\delta) + r^{(1)}_{\beta \delta}
\Theta(n-\beta)] (1 - \delta_{\beta \delta}) \nonumber \\
& = & \epsilon_\alpha \delta_{\alpha \gamma} + \sum_{\delta}
v^{(0)}_{\alpha \delta; \gamma \delta} \Theta(n-\delta) + \sum_{\beta,
  \delta} (v^{(0)}_{\alpha \beta; \gamma \delta} + v^{(0)}_{\gamma
  \beta; \alpha \delta}) r^{(1)}_{\delta \beta} \Theta(n-\delta)
\Theta(\beta-n-1) \;.
\end{eqnarray}
For $r^{(2)}$ this yields
\begin{eqnarray}
\label{r2}
r^{(2)}_{\alpha \mu} = {1\over \epsilon_{\alpha} -\epsilon_{\mu}} 
\sum_{\delta} \Theta(n-\delta) \left[ \sum_{\eta} (v^{(0)}_{\alpha 
\eta; \mu \delta} + v^{(0)}_{\mu \eta; \alpha \delta}) r^{(1)}_{\delta
\eta} \Theta(\eta-n-1) 
 + \sum_{\nu \neq \alpha} v^{(0)}_{\mu \delta; \nu \delta} 
r^{(1)}_{\alpha\nu} - v^{(0)}_{\alpha \delta; \alpha \delta}
r^{(1)}_{\alpha\mu} \right] \;.
\end{eqnarray}

For the diagonal matrix elements this yields
\begin{eqnarray}
v_{\alpha \beta; \alpha \beta} & = &\left(1 -\sum_{\mu \neq \alpha} 
({r_{\alpha \mu}^{(1)}})^2 - \sum_{\mu \neq \beta} ({r_{\beta
    \mu}^{(1)}})^2 \right) v^{(0)}_{\alpha \beta; \alpha \beta} +
\left[ 2\sum_{\mu \neq \alpha} (r^{(1)}_{\alpha  \mu} +
  r^{(2)}_{\alpha \mu}) v^{(0)}_{\mu \beta; \alpha \beta} + \alpha
  \leftrightarrow \beta \right] \nonumber \\ && + 2 \sum_{\mu \neq
    \alpha; \nu \neq \beta} r^{(1)}_{\alpha \mu} r^{(1)}_{\beta \nu}
  \left( v^{(0)}_{\mu \nu; \alpha \beta} + v^{(0)}_{\mu \beta; \alpha
      \nu} \right) + \left[ \sum_{\mu, \nu \neq \alpha}
    r^{(1)}_{\alpha \mu} r^{(1)}_{\alpha \nu} v^{(0)}_{\mu \beta; \nu
      \beta} + \alpha \leftrightarrow \beta \right] \;,
\end{eqnarray}
and, to fourth order in $v^{(0)}$
\begin{eqnarray}
\label{v-v-HF}
v_{\alpha \beta; \alpha\beta}\; v_{\alpha \beta; \alpha \beta} & = &
v^{(0)}_{\alpha \beta; \alpha \beta} v^{(0)}_{\alpha \beta; \alpha
  \beta} - 2 \left[ \sum_{\mu \neq \alpha} ({r_{\alpha\mu}^{(1)}})^2
v^{(0)}_{\alpha \beta; \alpha \beta} v^{(0)}_{\alpha \beta; \alpha
  \beta} + \alpha \leftrightarrow \beta \right] \nonumber \\
&& + \left[ 4  \sum_{\mu \neq \alpha} v^{(0)}_{\alpha \beta; \alpha
  \beta} r_{\alpha\mu}^{(2)} v^{(0)}_{\mu \beta; \alpha \beta} +
\alpha \leftrightarrow \beta \right] + \left[ 2 \sum_{\mu, \nu \neq
  \alpha} v^{(0)}_{\alpha \beta; \alpha \beta} r^{(1)}_{\alpha \mu}
r^{(1)}_{\alpha \nu} v^{(0)}_{\mu \beta; \nu \beta} + \alpha
\leftrightarrow \beta \right] \nonumber \\
&& + 4 \sum_{\mu \neq \alpha; \nu \neq \beta} v^{(0)}_{\alpha \beta;
  \alpha \beta} r^{(1)}_{\alpha \mu} r^{(1)}_{\beta \nu} \left(
v^{(0)}_{\mu \nu; \alpha \beta} + v^{(0)}_{\mu \beta; \alpha \nu}
\right) + \left[ 4 \sum_{\mu, \nu \neq \alpha} r^{(1)}_{\alpha \mu}
v^{(0)}_{\mu \beta; \alpha \beta} r^{(1)}_{\alpha \nu} v^{(0)}_{\nu
  \beta; \alpha \beta} + \alpha \leftrightarrow \beta \right]
\nonumber \\
&& + 8 \sum_{\mu \neq \alpha; \nu \neq \beta} r^{(1)}_{\alpha \mu} 
v^{(0)}_{\mu \beta; \alpha \beta} r^{(1)}_{\beta \nu} v^{(0)}_{\alpha
  \nu; \alpha \beta} \;.
\end{eqnarray}

\subsubsection{Average Values}

We first calculate the average HF matrix element in the ensemble
(iii), the two--body Gaussian ensemble. For a diagonal HF matrix
element $v_{\alpha \beta; \alpha \beta}$ we can rewrite
Eq.~(\ref{HF-element-g}) as
\begin{equation}
\label{HF-element}
v_{\alpha \beta; \alpha \beta} = v^{(0)}_{\alpha \beta; \alpha \beta}
+ 2 \sum_{\mu \neq \alpha} r^{(1)}_{\alpha \mu} v^{(0)}_{\mu \beta;
  \alpha \beta} + 2 \sum_{\mu \neq \beta} r^{(1)}_{\beta \mu}
v^{(0)}_{\mu \alpha; \beta \alpha} \;,
\end{equation}
where the third term on the r.h.s. is obtained from the second term by
exchanging $\alpha$ and $\beta$. To find the mean value of $v_{\alpha
\beta; \alpha \beta}$ from Eq.~(\ref{HF-element}), we calculate for
$\mu \neq \alpha$ 
\begin{eqnarray}
\label{v0-v0}
\overline{ v^{(0)}_{\mu \delta; \alpha \delta}
           v^{(0)}_{\mu \beta;  \alpha \beta}}
& = & 2u^2(\delta_{\delta \beta} - \delta_{\delta \beta} \delta_{\alpha
\beta} \delta_{\delta \alpha} 
+ \delta_{\mu \beta} \delta_{\delta \mu} \delta_{\alpha \beta}
\delta_{\delta \alpha} 
- \delta_{\mu \beta} \delta_{\delta \mu} \delta_{\delta \beta}
\nonumber \\
& & \;\;\;\;\;
            +\delta_{\mu \alpha} \delta_{\delta \beta}
- \delta_{\mu \beta} \delta_{\delta \alpha} \delta_{\alpha \mu}
+ \delta_{\mu \beta} \delta_{\delta \alpha} \delta_{\alpha \beta}
\delta_{\delta \mu}
- \delta_{\mu \alpha} \delta_{\delta \beta} \delta_{\alpha \beta}
\delta_{\delta \mu}) \nonumber \\
& = & 2u^2( \delta_{\delta \beta}- \delta_{\delta \beta}
           \delta_{\alpha \beta} - \delta_{\mu \beta} \delta_{\delta
           \mu} \delta_{\delta \beta}) = 2u^2 \delta_{\delta \beta}(1
           - \delta_{\alpha \beta} - \delta_{\mu \beta}) \;, 
\end{eqnarray}
where all other terms vanish since $\mu \neq \alpha$. Averaging
$v_{\alpha \beta; \alpha \beta}$ in Eq.~(\ref{HF-element}) using
Eq.~(\ref{v0-v0}), we find
\begin{eqnarray}
\label{HF-element1}
\bar v_{\alpha \beta; \alpha \beta} & = &\bar v^{(0)}_{\alpha \beta;
  \alpha \beta} + 4 u^2 (1-\delta_{\alpha \beta}) \left[
  \Theta(n-\beta) \left( \sum_{\mu \neq \alpha} {1\over
  \epsilon_\alpha - \epsilon_\mu} - {1 \over \epsilon_\alpha -
  \epsilon_\beta}\right) + \alpha \leftrightarrow \beta \right]
  \nonumber \\ 
& = & 4 u^2 (1-\delta_{\alpha \beta}) \left[ \Theta(n-\alpha)
  \sum_{\mu \neq \alpha, \beta} {1\over \epsilon_\beta - \epsilon_\mu}
  + \Theta(n-\beta) \sum_{\mu \neq \alpha, \beta} {1 \over
  \epsilon_\alpha - \epsilon_\mu} \right] \;.
\end{eqnarray}

For a general matrix element $v_{\alpha \beta; \gamma \delta}$ we find
\begin{equation}
\label{HF-element-av}
\bar v_{\alpha \beta; \gamma \delta} = \left(\sum_{\mu \neq \alpha} 
\overline{r^{(1)}_{\alpha \mu}v^{(0)}_{\mu \beta; \gamma \delta}}
\right) +(\alpha \leftrightarrow \beta, \gamma \leftrightarrow \delta) 
+(\alpha \leftrightarrow \gamma, \beta \leftrightarrow \delta) 
+(\alpha \leftrightarrow \delta, \beta \leftrightarrow \gamma)\;,
\end{equation}
where we have used Eq.~(\ref{HF-element-g}) and the symmetries of $v$. 
We have
\begin{eqnarray}
\label{r1-v}
\sum_{\mu \neq \alpha} \overline{r^{(1)}_{\alpha \mu} v^{(0)}_{\mu 
\beta; \gamma \delta}} & = & {1 \over \epsilon_\alpha -\epsilon_\mu}
\sum_\sigma \Theta(n-\sigma) \overline{v^{(0)}_{\mu \sigma; \alpha
    \sigma} v^{(0)}_{\mu \beta; \gamma \delta}} \;, \nonumber \\
\overline{v^{(0)}_{\mu \sigma; \alpha \sigma} v^{(0)}_{\mu \beta;
    \gamma \delta}} & = &  2 u^2 \delta_{\sigma \beta} (1 -
\delta_{\mu \beta}) (\delta_{\alpha \gamma} \delta_{\sigma \delta} -
\delta_{\alpha \delta} \delta_{\sigma \gamma}) \;.
\end{eqnarray}
Using Eqs.~(\ref{HF-element-av}) and (\ref{r1-v}), we find
\begin{equation}
\bar v_{\alpha \beta; \gamma \delta} = (\delta_{\alpha \gamma} 
\delta_{\beta \delta} - \delta_{\alpha \delta} \delta_{\beta \gamma})
\bar v_{\alpha \beta; \alpha \beta} \;,
\end{equation}
where $\bar v_{\alpha \beta; \alpha \beta}$ is given by
Eq.~(\ref{HF-element1}).

A similar calculation for the orthogonal ensemble~(ii) yields for the
average HF matrix element
\begin{eqnarray}
\label{HF-element2}
\bar v_{\alpha \beta; \alpha \beta} = 6 u^2 
(1 - \delta_{\alpha \beta}) \left[ \Theta(n-\alpha) \sum_{\mu \neq
  \alpha, \beta} {1 \over \epsilon_\beta - \epsilon_\mu} +
\Theta(n-\beta) \sum_{\mu \neq \alpha,\beta} {1 \over \epsilon_\alpha
  - \epsilon_\mu} \right] \;.
\end{eqnarray}
%\end{widetext}
Comparing this with Eq.~(\ref{HF-element1}) we see that the result is
the same as for the two--body Gaussian ensemble (iii) except for an
overall factor of $3/2$.

\subsubsection{Variance}

The expectation values of each of the seven terms on the r.h.s. of
Eq.~(\ref{v-v-HF}) can be calculated using Wick's theorem and
Eq.~(\ref{corr-goe-a}). The calculation proceeds along lines similar
to those of the previous Section. We do not give any details of this
rather lengthy calculation and confine ourselves to the result. For
the variance of the diagonal HF matrix elements we find
%\begin{widetext}
\begin{equation}\label{HF-variance}
\sigma^2(v_{\alpha \beta; \alpha \beta}) = 4 u^2 (1-\delta_{\alpha
  \beta}) \left[1 - 4 u^2 \Theta(n-\alpha) \sum_{\mu \neq \alpha,
  \beta} {1 \over (\epsilon_\beta - \epsilon_\mu)^2} - 4 u^2
  \Theta(n-\beta) \sum_{\mu \neq \alpha, \beta} {1 \over
  (\epsilon_\alpha  - \epsilon_\mu)^2} \right] \ .
\end{equation}
\end{widetext}

We observe that in contrast to the expressions derived for the average
values of the matrix elements, Eq.~(\ref{HF-variance}) cannot be
averaged over the GOE spectrum. This is because the integrals over the
eigenvalues possess a logarithmic singularity. We shall return to 
this point when we compare the numerical results with the perturbative
expressions in Section~\ref{variance1}.

\subsection{Statistics of the Hartree-Fock matrix elements: numerical
  results}\label{HF-elements}

We now report on numerical results concerning the statistics of the
interaction matrix elements in the self--consistent HF basis, using 
the true random--matrix ensembles of Section~\ref{ran}. We focus
attention on the diagonal matrix elements $v_{\sigma \tau; \sigma
\tau}$. Unless otherwise stated, we use in this Section the orthogonal
``non-local'' two--body ensemble (\ref{corr-goe-a}) with $m = 30$
single--particle states and $n=10$ electrons.

In the simulations, we construct the ensemble (\ref{eigen-ensemble})
as follows. We generate an $m\times m$ GOE matrix $h^{(0)}$ and
diagonalize it to find its spectrum $\epsilon_\alpha$. We also
generate Gaussian variables $v_{\alpha\beta;\gamma\delta}$ that are
uncorrelated with the $\epsilon_\alpha$'s and satisfy
Eq.~(\ref{corr-goe}). We then compute the corresponding
antisymmetrized interaction matrix elements from $v^A_{\alpha \beta;
  \gamma \delta} = v_{\alpha \beta; \gamma \delta} - v_{\alpha \beta;
  \delta \gamma}$.

The parameters of the ensemble are $\Delta$ and $u$. We have chosen
the variance of the off--diagonal matrix elements of the GOE matrix 
$h^{(0)}$ to be $\sigma^2(h^{(0)}_{\alpha\beta})= 3 m / 4 \pi^2$,
so that the mean level spacing in the middle of the spectrum 
is fixed at $\Delta = \sqrt{3}/2 \approx 0.866$. For each of the
quantities presented in the figures of this Section we have used
$10,000$ realizations.

In some of the calculations presented below, the single--particle
spectrum is kept fixed and an ensemble of antisymmetrized two--body
interaction matrix elements $v^A_{\alpha \beta; \gamma \delta}$ is
generated as discussed above. 

We used a rather low--dimensional GOE matrix to generate the one--body
ensemble. An alternative choice would have been to use the center
section of a high--dimensional GOE matrix as our starting point. Our
choice has the obvious disadvantage of being not free of edge effects. 
However, the comparison made below of some of our results with those
obtained from a picket--fence spectrum (constant spacings) shows that
edge effects are small for the quantities we investigate. Our choice
has the obvious advantage of keeping the dimension of all relevant
matrices small. This is very useful in view of the number of
realizations we have chosen, and of the need to perform a HF iteration
for each of these realizations.

\subsubsection{Distributions}

We first consider the case of a fixed single--particle spectrum,
allowing only the single--particle wave functions and, thus, the
two--body matrix elements to vary randomly. For each realization of
the two--body matrix elements, we solve the HF equations and compute
the diagonal matrix elements $v_{\alpha \beta; \alpha \beta}$ in the
HF single--particle basis. 

\begin{figure}[ht!]
\vspace{5 mm}
\centerline{\includegraphics[scale=1.0]{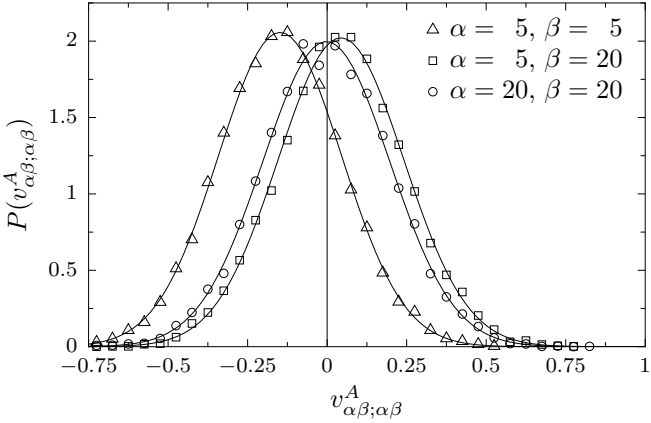}}
\vspace{3 mm}
\caption {The distributions $P(v_{\alpha \beta; \alpha \beta})$ of
  three diagonal matrix elements in the HF basis for a fixed
  single--particle GOE spectrum in the ensemble (\ref{corr-goe-a})
  with $u = 0.1$. There are $n = 10$ electrons. The symbols denote
  the results of the numerical simulations. The solid lines are
  Gaussian fits.}
 \label{dist}
\end{figure}

Fig.~\ref{dist} shows the distributions of three diagonal matrix
elements $v_{\alpha \beta; \alpha \beta}$ for $n = 10$ electrons
and $u=0.1$.
For $\alpha=5, \beta=5$ both levels are filled, for $\alpha=5,
\beta=20$ one level is filled, the other is empty, and for $\alpha=20,
\beta=20$, both levels are empty. The distributions are all well
described by Gaussians (solid lines). The same holds true for the
distributions generated by choosing both single--particle energies
and two--body matrix elements at random. It is, thus, sufficient in
all cases to consider only the first two moments of $v_{\alpha \beta;
  \alpha \beta}$.

\subsubsection{Average HF matrix elements}

\begin{figure}[ht!]
\vspace{5 mm}
\centerline{\includegraphics[scale=0.92]{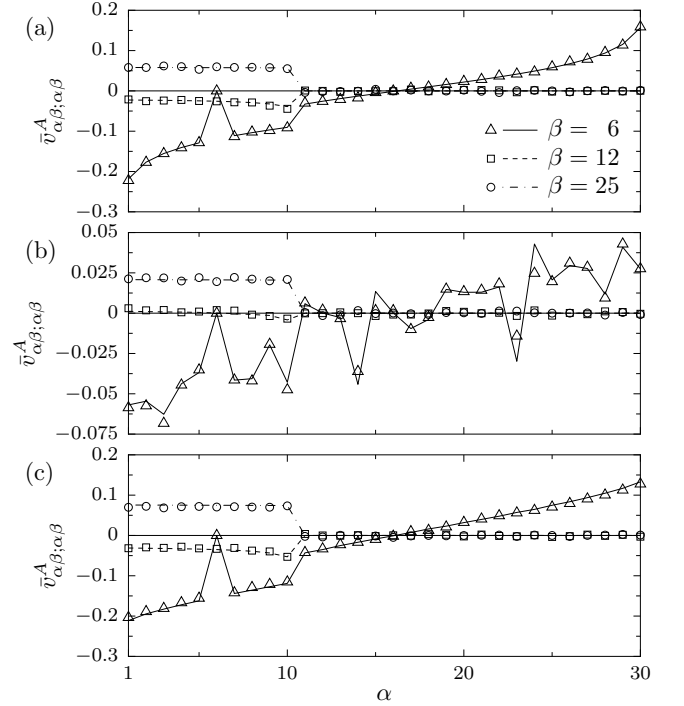}}
\caption {Average matrix elements $\bar v_{\alpha \beta; \alpha
    \beta}$ in the HF basis versus $\alpha$ for fixed $\beta$.  We show
  results for three values of $\beta$ ($\beta=6, 12, 25$). We use the
  ensemble~(\ref{corr-goe-a}) with $m = 30$ and for $n = 10$
  electrons. (a) The data are generated for $u= 0.1$ and for a fixed
  picket--fence single--particle spectrum (equal spacings). The
  symbols give the numerical simulations, and the lines give the
  perturbative result~(\ref{HF-element1}). (b) As in panel (a) but for
  $u = 0.05$ and for a fixed single--particle GOE spectrum. (c) The
  results are for the same ensemble as in (a), but the average is
  taken over the full ensemble~(\ref{corr-goe-a}), i.e., over both the
  single--particle GOE spectrum and the two--body interaction matrix
  elements of Eq.~(\ref{eigen-ensemble}). The symbols represent the
  numerical simulations and the lines the perturbative
  expression~(\ref{HF-element1}), averaged over the single--particle
  GOE spectrum.}
\label{av}
\end{figure}

The top two panels (a) and (b) of Fig.~\ref{av} show the average
values of some diagonal matrix elements calculated for a fixed
single--particle spectrum. This fixed spectrum was chosen as a
picket--fence spectrum (equal spacings) in panel (a) of Fig.~\ref{av}
and as a GOE spectrum in panel (b) of Fig.~\ref{av}. The perturbative
expression~(\ref{HF-element1}) shown by solid, dashed and
dashed--dotted lines is in good agreement with the numerical results.
The agreement is better for a picket--fence spectrum than for a GOE
spectrum.

We turn to the general case where the statistics are collected by
varying both the single--particle GOE spectrum and the
single--particle wave functions. Such fully averaged HF matrix
elements are shown in panel (c) of Fig.~\ref{av}.

\begin{figure}[ht!]
\vspace{5 mm}
\centerline{\includegraphics[scale=0.95]{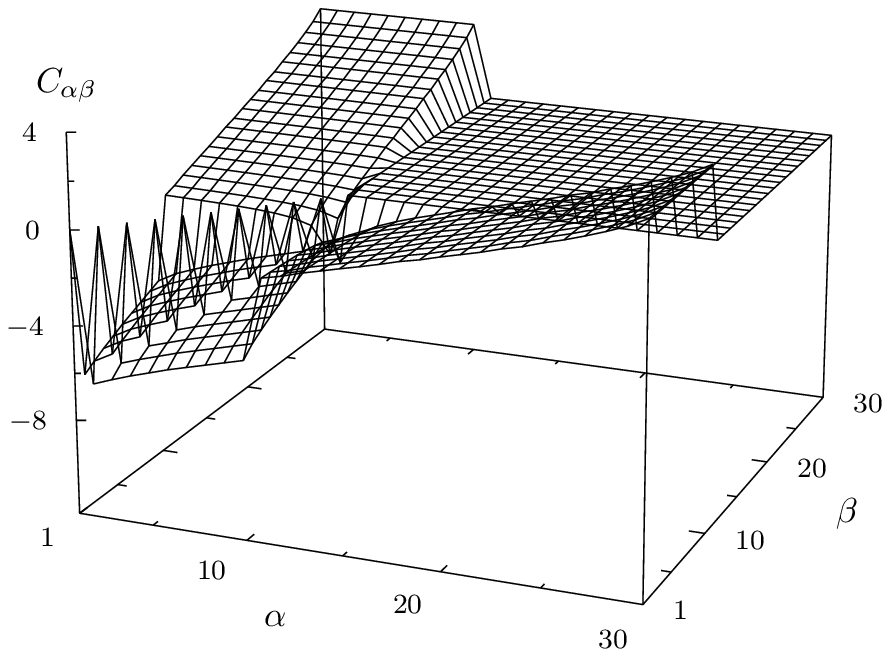}}
\vspace{3 mm}
\caption {Three--dimensional plot of $C_{\alpha \beta}$ [see
  Eq.~(\ref{c-pert})] versus $\alpha$ and $\beta$. The results shown
  are for $n = 10$ electrons occupying $m = 30$ orbitals.}
\label{3dpert}
\end{figure}

According to the perturbative expression~(\ref{HF-element1}), the full
ensemble average of a HF diagonal matrix element is given by $4u^2
(1-\delta_{\alpha \beta}) C_{\alpha\beta}$  with
%\begin{widetext}
\begin{eqnarray}\label{c-pert}
C_{\alpha\beta} &= & \left[ \Theta(n-\alpha)  \left\langle\sum_{\mu \neq
\alpha, \beta} {1\over \epsilon_\beta - \epsilon_\mu}\right\rangle \right. \nonumber
 \\ && \;\;\; +   \left.
\Theta(n-\beta) \left\langle\sum_{\mu \neq \alpha, \beta} {1 \over
  \epsilon_\alpha - \epsilon_\mu}\right\rangle \right] \;.
\end{eqnarray}
%\end{widetext}
The symbol $\langle \ldots\rangle$ denotes the GOE average. In
Fig.~\ref{3dpert} we show a three--dimensional plot of $C_{\alpha
\beta}$ versus $\alpha$ and $\beta$ for $m = 30$ and $n = 10$.

\begin{figure}[ht!]
\vspace{5 mm}
\centerline{\includegraphics[scale=0.95]{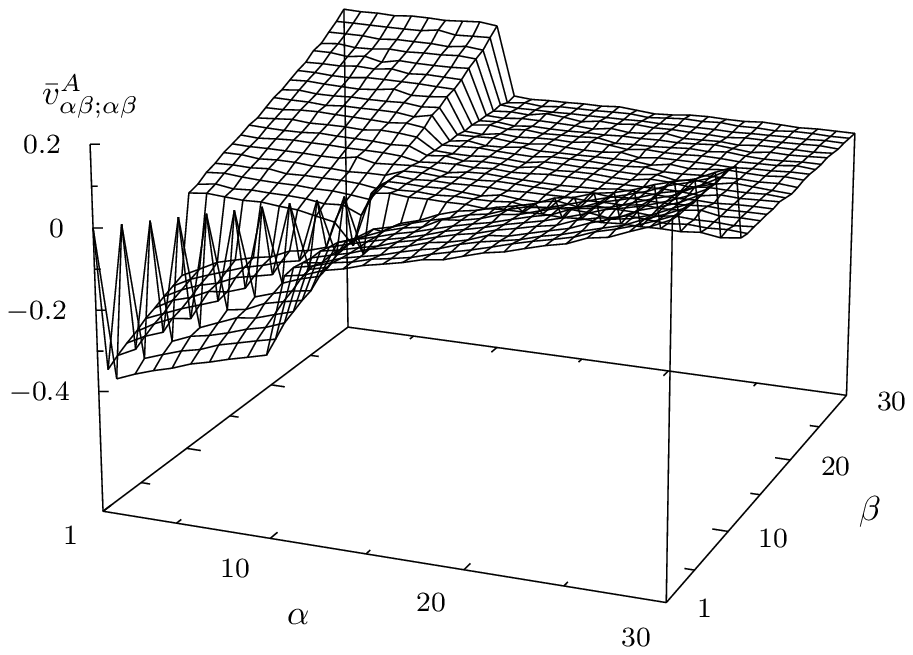}}
\vspace{3 mm}
\caption {Three--dimensional plot of the full ensemble average of
  $v_{\alpha \beta; \alpha \beta}$ versus $\alpha$ and $\beta$. We use
  the ``local'' interaction ensemble with $u = 0.1$ for $n = 10$
  electrons occupying $m = 30$ orbitals. The results shown here agree
  well with the perturbative expression $6u^2 C_{\alpha\beta}$ where
  $C_{\alpha\beta}$ is shown in Fig.~\ref{3dpert}.}
\label{3dnum}
\end{figure}

For comparison, we show in Fig.~\ref{3dnum} the full ensemble average
of a diagonal HF matrix element (averaged numerically over both the
two--body matrix elements and the single--particle GOE spectrum) versus
$\alpha$ and $\beta$ for $u=0.1$. The result agrees well with $6 u^2
C_{\alpha \beta}$ if we take $C_{\alpha \beta}$ from Fig.~\ref{3dpert}.

\subsubsection{Variance}
\label{variance1}

\begin{figure}[ht]
\vspace{5 mm}
\centerline{\includegraphics[scale=0.97]{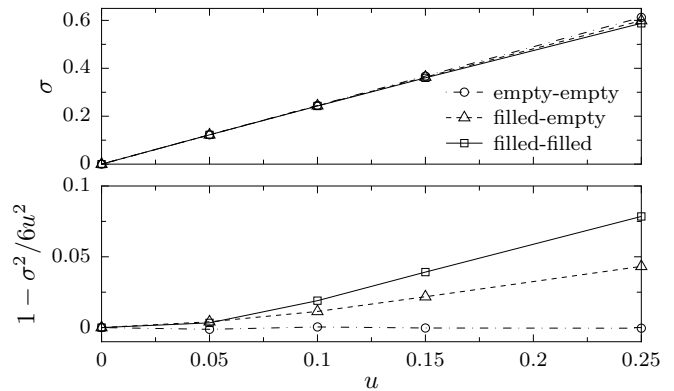}}
\vspace{3 mm}
\caption {Top panel: The standard deviation of a HF matrix element
  $\sigma( v_{\alpha \beta; \alpha \beta})$ versus $u$ for the
  ``local'' interaction ensemble~(\ref{corr-local-a}). We show results
  for three different types of diagonal matrix elements, corresponding
  to filled--filled, filled--empty and empty--empty single--particle
  levels. The results shown for each type are averaged over all matrix
  elements of that type. Bottom panel: The fractional deviation $1 -
  \sigma^2 / 6u^2$ of the variance $\sigma^2$ of the matrix elements
  in the HF basis from their variance $6 u^2$ in the non--interacting
  basis versus $u$. Same conventions as in the top panel.}
\label{sigma}
\end{figure}

The two--body matrix elements which serve as input in the HF
calculation have a Gaussian distribution with second moments defined
in Section~\ref{ortho}. Our numerical simulations indicate that the
variances $\sigma^2$ of the HF matrix elements are very close to
these input values. We observe that in the HF basis $\sigma$ decreases
slightly, the size of the decrease depending on whether the diagonal
matrix element corresponds to filled--filled, filled--empty or
empty--empty single--particle orbitals. There are fluctuations because
of the finite size of the ensemble. To improve the statistics we
further average the variances separately for these three different
types of matrix elements. The top panel of Fig.~\ref{sigma} shows the 
resulting standard deviation $\sigma(v_{\alpha \beta; \alpha \beta})$
for the ``local'' interaction ensemble~(\ref{corr-local-a}) versus $u$.
The input value is $\sigma = \sqrt{6} u$. The difference between the
HF results and the input value is amplified in the bottom panel of
Fig.~\ref{sigma}. The variance of the empty--empty matrix elements
is practically unchanged, while the largest decrease is observed for
the filled--filled matrix elements.

We have also compared our numerical results for a fixed choice of the
single--particle spectrum with the perturbative
expression~(\ref{HF-variance}). This expression seems to fail no
matter how small $u$ was chosen. (When $u$ becomes too small, the
statistical noise prohibits a meaningful comparison). While the
numerical result is always close to $4 u^2$, this is not true for the
expression~(\ref{HF-variance}) which at times may even become negative.
We speculate that this fact may signal a breakdown of perturbation
theory but have not followed up this point.

\subsection{Statistics of the Hartree--Fock single--particle
 energies}
\label{HF-spect}

The statistical results presented in the following are calculated 
for the entire ensemble, i.e., by collecting statistics over both
the single--particle spectrum and the single--particle wave functions
(i.e., the interaction matrix elements). Unless stated otherwise, the
number of realizations used in the numerical simulations is $10,000$
(for each value of $u$).

\begin{figure}[ht]
\vspace{5 mm}
\centerline{\includegraphics[scale=1.0]{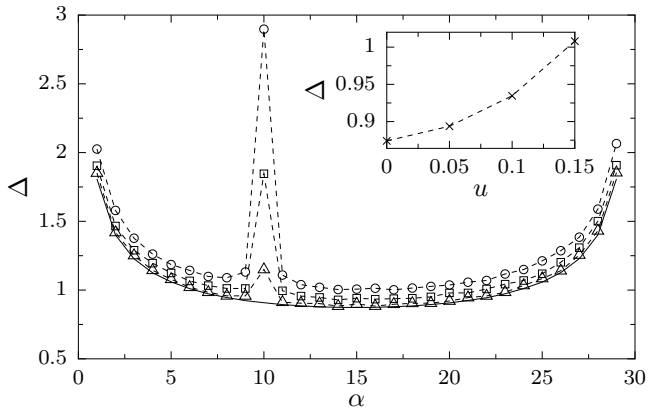}}
\vspace{3 mm}
\caption {The mean level spacing $\Delta_{\rm HF}$ of the
  single--particle HF levels versus level index $\alpha$ for $u =
  0.05$ (triangles), $u = 0.1$ (squares) and $u = 0.15$ (circles). The
  results shown are for $n = 10$ electrons and $m = 30$
  single--particle states. The solid line describes the mean level
  spacing $\Delta$ for a GOE of dimension $m = 30$. Inset: the mean
  level spacing in the center of the spectrum versus $u$.}
\label{delta}
\end{figure}

The mean level spacing $\Delta$ of the single--particle GOE levels is
given approximately by the inverse semi--circle law and increases
towards the edges of the spectrum. We now consider the mean level
spacing $\Delta_{\rm HF}$ of the HF single--particle spectrum
calculated by averaging over the entire ensemble. Not surprisingly, we
find that $\Delta_{\rm HF}$ increases monotonically with $u$.
Fig.~\ref{delta} shows $\Delta_{\rm HF}$ versus the level index
$\alpha$ for different values of $u$. The solid line is the GOE
result. We notice the large peak at $n = 10$. It is due to a large
spacing between the highest filled and the lowest empty HF level. This
is a well--known feature of HF calculations and is referred to as the
HF gap. We notice also the slight increase in the mean spacing close
to the HF gap. In our discussion of the statistics of the HF
single--particle energies $\epsilon_\alpha^{(n)}$, we treat filled
levels, empty levels and the HF gap separately.

\subsubsection{Filled levels and empty levels}
\label{nearest-neighbor}

\begin{figure}[ht]
\vspace{5 mm}
\centerline{\includegraphics[scale=1.0]{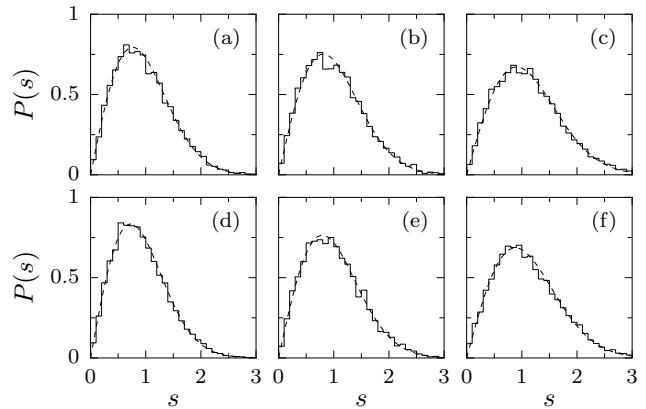}}
\vspace{3 mm}
\caption {Nearest--neighbor spacing distribution $P(s)$ versus
  spacing $s$ where $s$ is measured in units of the local mean level
  spacing of the corresponding HF levels. Top panel: The histograms
  show $P(s)$ for the highest two filled single--particle HF levels
  $\epsilon^{(n)}_{n-1}$ and $\epsilon^{(n)}_{n}$ (with $s =
  (\epsilon^{(n)}_{n} - \epsilon^{(n)}_{n-1})/ \langle
  \epsilon^{(n)}_{n} - \epsilon^{(n)}_{n-1}\rangle$), and for
  three values of $u$: (a) $u = 0.05$; (b) $u = 0.1$; and (c)
  $u = 0.15$. The dashed lines give the GOE Wigner distribution. 
  Bottom panels: same as in the top panels but for the two lowest
  empty single--particle HF levels $\epsilon^{(n)}_{n+1}$ and
  $\epsilon^{(n)}_{n+2}$. The values of $u$ are (d) $u = 0.05$;
  (e) $u=0.1$; and (f) $u=0.15$.}
\label{filledemptypofs}
\end{figure}

Let $s$ denote the spacing of a pair of successive filled or empty
levels, measured in units of the local mean level spacing. The
nearest--neighbor spacing distribution $P(s)$ is found to follow
the GOE Wigner distribution. We observe such behavior even for the
pair of filled (empty) levels closest to the HF gap. This is
demonstrated in Fig.~\ref{filledemptypofs} where the numerical
distributions (histograms) for $P(s)$ are compared with the Wigner
distribution (dashed lines) for different values of $u$ and for
both the highest two filled HF levels (top panels) and the lowest
two empty HF levels (bottom panels).

\subsubsection{Hartree--Fock gap}
\label{gap}

\begin{figure}[ht!]
\vspace{5 mm}
\centerline{\includegraphics[scale=1.0]{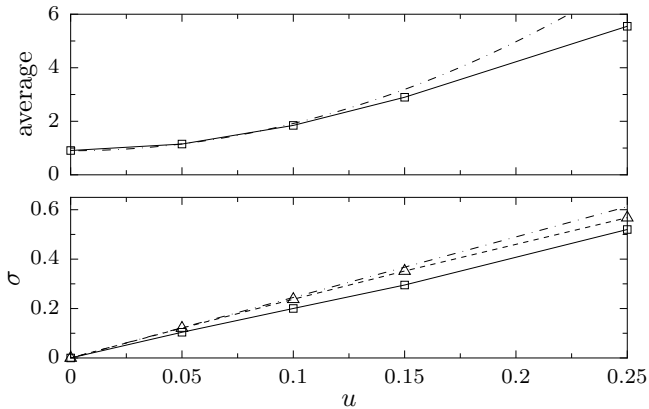}}
\vspace{3 mm}
\caption {Top panel: average HF gap versus $u$. The squares are the
  results of simulations of the full ensemble~(\ref{eigen-ensemble}).
  The dotted--dashed line is obtained by using Eq.~(\ref{HF-gap})
  and a spectral GOE average of the perturbative 
  expression~(\ref{HF-element1}).
  Bottom panel: the squares depict the width $\sigma$ of a Gaussian
  the convolution of which with a Wigner distribution is fitted to
  the gap distributions in Fig.~\ref{hfgap}. The dotted--dashed line
  is $\sqrt{6}u$, and the triangles (connected by the dashed line)
  are the standard deviations $\sigma(v_{n+1,n})$ of the
  corresponding HF interaction matrix element (see text).}
\label{hfgapparam}
\end{figure}

We simplify the notation by writing $v_{\alpha \beta} \equiv
v_{\alpha \beta;\alpha\beta}$. In the HF basis and for $n$ electrons,
the HF gap is given by
\begin{equation}\label{HF-gap}
\epsilon^{(n)}_{n+1} - \epsilon^{(n)}_n = (h^{(0)}_{n+1,n+1} -
h^{(0)}_{n, n}) + v_{n+1,n} +\sum_{\tau = 1}^{n-1} (v_{n+1,\tau} -
v_{n,\tau}) \;.
\end{equation}
The squares in the top panel of Fig.~\ref{hfgapparam} show the
numerical results for the average gap, calculated from the full
ensemble~(\ref{eigen-ensemble}) with a ``local'' interaction
[Eq.~(\ref{corr-local-a})]. To estimate the average gap, we use
in Eq.~(\ref{HF-gap}) the perturbative expression~(\ref{HF-element2})
for the diagonal HF matrix elements which we average over a
single--particle GOE spectrum (see $C_{\alpha\beta}$ in
Section~\ref{HF-elements}). We also need to know the average values
of $h^{(0)}_{n+1,n+1} - h^{(0)}_{n, n}$ which are found to be $0.895,
0.868, 0.836$, and $0.765$ for $u=0.05, 0.1, 0.15$ and $0.25$,
respectively (the corresponding value in the absence of the
two--body interaction is $\Delta=0.908$). The resulting estimate for
the average gap depends quadratically on $u$, and is shown by the
dotted--dashed line in the top panel of Fig.~\ref{hfgapparam}. For
small values of $u$ ($\alt 0.15$) our estimate provides a good
approximation for the average HF gap, but for larger values there
are deviations, indicating the breakdown of the lowest--order
perturbative approach. We note that the average interaction
(\ref{average-int}) contributes a constant $v_0$ to the 
average gap. In our calculations we have set the average 
interaction to zero, and this constant (but large) contribution 
to the average gap is not shown in our numerical results.

\begin{figure}[ht!]
\vspace{5 mm}
\centerline{\includegraphics[scale=1.0]{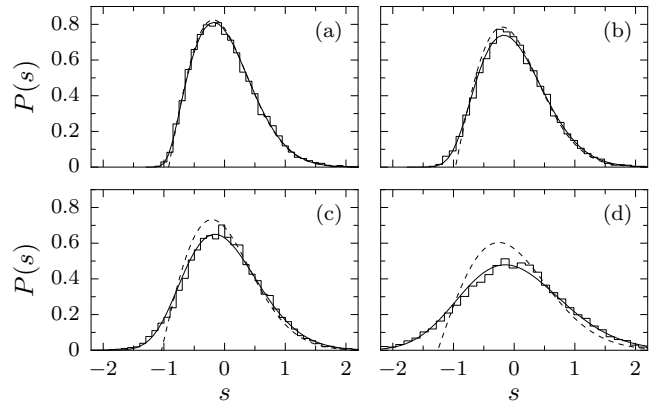}}
\vspace{3 mm}
\caption {The calculated distribution of the shifted HF gap
  (histograms) for four values of $u$: (a) $u = 0.05$; (b) $u = 0.1$;
  (c) $u = 0.15$; and (d) $u = 0.25$. The gap is shifted to have the
  average value zero. The solid lines describe convolutions of a
  shifted Wigner distribution (with an appropriate $\Delta$) with a
  Gaussian. The width of the Gaussian is fitted. The dashed lines are
  the shifted Wigner distributions.}
\label{hfgap}
\end{figure}

The distribution $P(s)$ of the HF gap, shifted by its average value,
is shown in Fig.~\ref{hfgap} for several values of $u$ (histograms).
In Section~\ref{peak-spacing}, we motivate an approximation to this
gap distribution as a convolution of a (shifted) Wigner distribution
with an appropriate $\Delta$, and a Gaussian. The dashed lines in 
Fig.~\ref{hfgap} are the corresponding Wigner distributions, while
the solid lines describe the convolutions for which the width
$\sigma$ of the Gaussian is fitted. This fitted value of $\sigma$ is
shown in the bottom panel of Fig.~\ref{hfgapparam} versus $u$
(squares). The dotted--dashed line in Fig.~\ref{hfgapparam} is the
input value $\sqrt{6}u$ for the standard deviation of a diagonal
matrix element. The triangles give the standard deviation
$\sigma(v_{n+1,n})$ of the diagonal matrix element $v_{n+1,n}$ in
the HF basis for $n$ electrons. We observe that $\sigma$ is somewhat
smaller than the standard deviation of the corresponding HF
interaction matrix element.

\subsubsection{Spacing correlator}

In Section~\ref{nearest-neighbor} we have seen that the
nearest--neighbor spacing distributions of both, the occupied and the
empty HF states, follow RMT statistics. But what about the correlation
between occupied and empty levels? In this Section we show that the HF
gap modifies that correlation so that it differs from the RMT
prediction. To that end, we define the spacing correlator
\begin{equation}
\label{spacing-corr}
c_\alpha = {\overline{s_\alpha s_{\alpha +2}} - \bar s_\alpha \bar
  s_{\alpha+2} \over \sigma(s_\alpha)\sigma(s_{\alpha+2})} \;, 
\end{equation}
where $s_\alpha = \epsilon^{(n)}_\alpha - \epsilon^{(n)}_{\alpha-1}$
($\alpha = 2, \ldots, m$) are nearest--neighbor spacings and
$\sigma(s_\alpha)$ is the standard deviation of $s_\alpha$. For the HF
ensemble, the correlator $c_n$  measures the spacing correlation
across the gap, i.e., the correlation between the highest spacing of
the filled levels and the lowest spacing of the empty levels.  

For the GOE, the correlator~(\ref{spacing-corr}) is denoted by $C(r=1;
t=0)$ and explicitly given in Ref.~\onlinecite{brody81}. The correlator is
negative and approximated by  $C(r=1,t=0) \approx -1/[4 \pi^2 (4/\pi
- 1)] \approx -0.093$. A more accurate numerical value is $-0.087$. 

In the HF approach, the correlator~(\ref{spacing-corr}) is small and
exhibits large sample--to--sample fluctuations. Thus, a large number
of realizations is required to obtain a reliable value. Table I shows
$c_n$ of the HF single--particle spectrum for several values of $u$.
The results were obtained using $50,000$ realizations (for each $u$)
of the ``local'' interaction ensemble with $m = 30$.  We observe that
the correlator is negative and its magnitude becomes smaller as $u$
increases. Hence the HF gap has the effect of weakening the correlation
between filled and empty levels.

\begin{table}[ht!]
\begin{tabular} {c|c}
 $u$ &  correlator $c_n$ \\
\hline
   0 & $-0.084 \pm 0.011$ \\
 0.05 & $-0.078 \pm 0.011$ \\
 0.10 & $-0.061 \pm 0.012$ \\
 0.15 &  $-0.048 \pm 0.011$\\
 0.25 &  $-0.027 \pm 0.013$ \\
\hline
\end{tabular}
\caption{\label{TableI} The spacing correlator $c_n$ across the HF gap
  for several values of $u$ (using $50,000$ realizations for each
  $u$).}
\end{table}

On the other hand, when both spacings are between filled states ($\alpha \leq
n-2$) or both between empty states ($\alpha \geq n+2$), we find that $c_\alpha$ is
close to its standard GOE value (except for edge effects). This provides
additional support for the hypothesis that the filled and the empty HF
states separately satisfy RMT statistics.

\begin{figure}[ht!]
\vspace{5 mm}
\centerline{\includegraphics[scale=0.98]{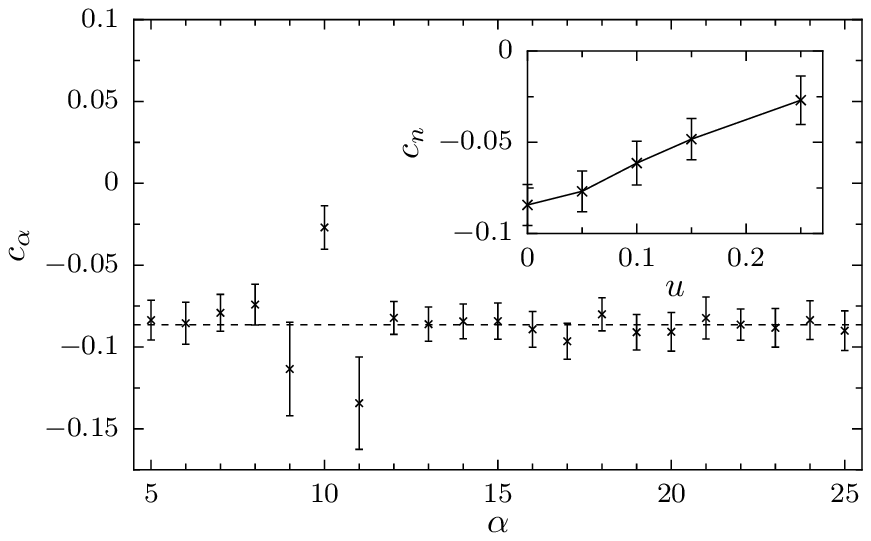}}
\vspace{3 mm}
\caption {The spacing correlator $c_\alpha$ [see Eq.~(\ref{spacing-corr})]
  versus $\alpha$ for the HF spectrum of the ``local'' interaction ensemble
  with $u = 0.25$ and $m = 30$ (values of $\alpha$ near the edges of the
  spectrum are omitted). The results shown are calculated for $50,000$
  realizations of the ensemble and include statistical errors. The
  dashed line is the GOE value of the correlator $-0.0863 \pm 0.0008$
  calculated from 500,000 realizations and averaged over
  $\alpha$. Inset: the correlator $c_n$ across the gap versus $u$.}
\label{s-correlator}
\end{figure}

Fig.~\ref{s-correlator} demonstrates the behavior of the correlator $c_\alpha$
as a function of $\alpha$ for $u=0.1$ and $u=0.25$. For $\alpha \leq n-2$ and $\alpha\geq n+2$, $c_\alpha$ fluctuates around its RMT value. However, the correlator
$c_n$ between spacings on opposite sides of the gap decreases in magnitude
with $u$, while the correlators $c_{n-1}$ and $c_{n+1}$ (for which one of
the spacings is the gap itself) are enhanced in magnitude compared with
the RMT value.

\subsection{Statistics of the Hartree-Fock single--particle
wave functions}
\label{HF-wavefunctions}

The induced ensemble is invariant under orthogonal (unitary)
transformations of the single--particle basis. We show now that the
corresponding single--particle HF ensemble is also invariant under
orthogonal (unitary) transformations. For simplicity, we present the
proof for the orthogonal case only.

We consider a particular realization~(\ref{ensemble-a}) of the
induced ensemble in a fixed basis $|i\rangle$. Under an orthogonal
transformation $\hat O$ of the single--particle space with
$a^{\dagger}_i \to \hat O a^\dagger_i \hat O = \sum_{i'} a^\dagger_{i'}
O_{i' i}$, we obtain a new realization of the many--body ensemble,
\begin{equation}
\label{eigen-ensemble-1}
H \to \tilde H = \hat O H \hat O^{-1} = \sum_{ij} \tilde h^{(0)}_{i j}
a^\dagger_i a_j + {1 \over 4} \sum_{ij;kl} \tilde v^A_{ij;kl}
a^\dagger_i a^\dagger_j  a_l a_k \;,
\end{equation}
where the new single--particle Hamiltonian $\tilde h^{(0)}$ and
interaction matrix elements $\tilde v^A$ are related to $h^{(0)}$
and $v^A$ by
\begin{eqnarray}
\tilde h^{(0)}_{i k} &  = &  \sum_{j l} O_{i j} h^{(0)}_{j l} (O^{-1})_{l
k}\;; \nonumber \\ \tilde v^A_{i j;k l} & = & \sum_{i'j'k'l'} O_{i i'} O_{j j'}
v^A_{i'j';k'l'} (O^{-1})_{k' k} (O^{-1})_{l' l} \;.
\end{eqnarray}
The Hamiltonians $h^{(0)}$ and ${\tilde h^{(0)}}$ have identical
single--particle spectra but different single--particle eigenstates.
Both are members of the single--particle GOE. Likewise, the
Hamiltonians $H$ and $\tilde H$ are both members of the induced
many--body ensemble.

In the fixed basis the single--particle HF Hamiltonian~(\ref{HF-Ham})
has the form
\begin{equation}
\label{fixed-HF-Ham}
h_{i k} =  h^{(0)}_{i k} + \sum_{jl} v^A_{ij;kp} \rho_{pj}\;.
\end{equation}
Here $\rho_{p j} = \langle j |\hat \rho |p \rangle = \sum_l \psi^*_l(p)
\psi_l(j)$ is the matrix representation of the single--particle density
operator $\hat\rho = \sum_{l = 1}^n |\psi_l \rangle \langle \psi_l |$.
As in Eq.~(\ref{density-matrix}), $| \psi_l \rangle$ denotes the
$l^{th}$ HF single--particle eigenstate, $\psi_l(j)$ its projection
onto the fixed basis state $j$, and the sum is over the lowest $n$ HF
levels. We consider the transformed single--particle
Hamiltonian
\begin{equation}
\label{tilde-h}
\tilde h_{i k} =  \sum_{j l} O_{i j} h_{j l} (O^{-1})_{l k} \ .
\end{equation}
To prove the orthogonal invariance of the ensemble of HF
single--particle Hamiltonians, we have to show that ${\tilde h}$ is
the HF Hamiltonian for the new realization~(\ref{eigen-ensemble-1}) of
the many--body ensemble. Using Eq.~(\ref{fixed-HF-Ham}) we can write
$\tilde h$ as
\begin{equation}
\label{new-HF-Ham}
\tilde h_{i k} =  \tilde h^{(0)}_{i k} + \sum_{jl} \tilde v^A_{ij;kl}
\tilde \rho_{lj} \ ,
\end{equation}  
where
\begin{equation}
\hat {\tilde \rho} = \hat O \rho \hat O^{-1} = \sum_{l=1}^n |\tilde
\psi_l \rangle \langle \tilde\psi_l| \ , \ {\rm with} \ | \tilde
\psi_l \rangle = {\hat O} | \psi_l \rangle \ .
\label{c1}
\end{equation}
This shows that $\tilde h_{i k}$ is indeed the HF Hamiltonian of the
realization~(\ref{eigen-ensemble-1}) provided that $\hat {\tilde
\rho}$ is the density matrix of the lowest $n$ eigenstates of
$\tilde h$. But Eq.~(\ref{tilde-h}) shows that $\tilde h$ has the same
eigenvalues $\epsilon^{(n)}_l$ as $h$ and that the eigenvectors of
$\tilde h$ are given by the second of Eqs.~(\ref{c1}). This completes
the proof. We conclude that the invariance of the many--body ensemble
under orthogonal transformations implies the orthogonal invariance of
the HF ensemble. In the GOE, the statistics of the eigenvectors are
determined by the orthogonal invariance of the ensemble. Thus we expect
the eigenvectors of the HF ensemble to obey the same statistics. For
example, the distribution of the components $\psi_l(j)$ of the HF
eigenvector $| \psi_l \rangle$ in the fixed basis of states $| j
\rangle$ must depend on orthogonal invariants only. For a single
eigenvector the only such orthogonal invariant is $\sum_j \psi^2_l(j)$. Since the eigenvector is normalized to unity, the probability
density is
\begin{equation}
\label{dist-vector}
P(\psi_l(1),\psi_l(2),\ldots) \propto \delta(\sum_j \psi^2_l(j) - 1)
\;.
\end{equation}
As for the GOE, integrating over all components except the $j$-th, we
find
\begin{equation}
\label{dist-component}
P(\psi_l(j)) = \pi^{-1/2} { \Gamma(\frac{m}{2})\over
\Gamma(\frac{m- 1}{2})} [1 - \psi^2_l(j)]^{m-3\over 2} \;,
\end{equation}
where $m$ is the dimension of the single--particle space. In the limit
of large $m$, $P(\psi_l(j))$ can be approximated by a Gaussian,
$P(\psi_l(j)) \propto e^{-m \psi^2_l(j)/2}$.  

We have computed numerically the statistics of an HF eigenfunction
component in the ``local'' interaction orthogonal
ensemble~(\ref{corr-local-a}). Since there is an ambiguity of the
overall sign of the wave function, we have calculated the distribution
of $y=m \psi^2_l(j)$ (the normalization is chosen to satisfy $\bar
y = 1$). It follows from (\ref{dist-component}) that
\begin{equation}\label{y-dist}
P(y) = \pi^{-1/2} { \Gamma(\frac{m}{2})\over \Gamma(\frac{m- 1}{2})}
(my)^{-1/2} \left(1 -{y\over m}\right)^{m-3\over 2} \;.
\end{equation}
In the limit of large $m$, this is just the Porter--Thomas distribution
\begin{equation}\label{Porter-Thomas}
P(y) = (2\pi y)^{-1/2} e^{-y/2} \;.
\end{equation}

\begin{figure}[ht]
\vspace{5 mm}
\centerline{\includegraphics[scale=1.0]{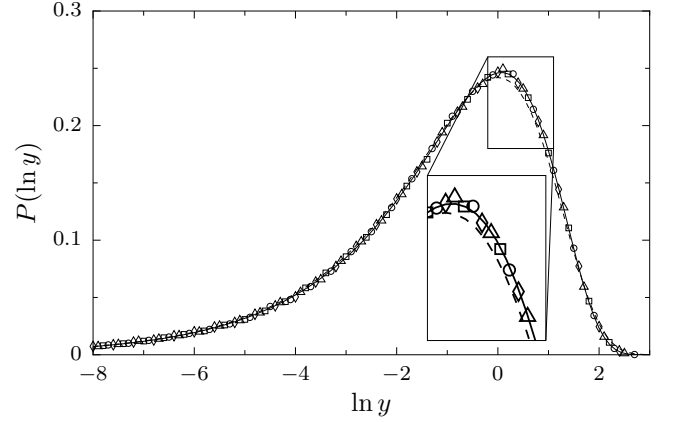}}
\vspace{3 mm}
\caption {The distributions $P(\ln y)$ versus $\ln y$, where $y$ is the
  square of an HF wave function component (scaled to give $\bar y =
  1$) for $u = 0.05$ (triangles), $u=0.1$ (squares), $u=0.15$
  (circles) and $u=0.25$ (diamonds). We use the ``local'' interaction
  orthogonal ensemble~(\ref{corr-local-a}) with $m = 30$ and $n=10$.
  The solid line  is the finite--$m$ RMT distribution~(\ref{y-dist})
  and the dashed line is the Porter--Thomas
  distribution~(\ref{Porter-Thomas}). A section of the graph is
  magnified to render more details.}
\label{psi2}
\end{figure}

Fig. \ref{psi2} shows numerical results for the distribution of
$P(\ln y)$ versus $\ln y$ for an HF eigenstate in the middle of the
spectrum and for $m = 30$. Since the distribution is independent of
the particular component, we collect statistics from all components.
Results are shown for $u=0.05, 0.1, 0.15$ and $0.25$ (symbols). For
all values of $u$ the distributions are well described by
Eq.~(\ref{y-dist}) (solid line). For reference we also show the
Porter--Thomas distribution (dashed line). The observed small
deviation from the limiting case of a Porter--Thomas distribution
is a finite--size effect.

As in the GOE, it is also possible to calculate from
Eq.~(\ref{dist-vector}) the correlator for two different squared
components of an HF eigenfunction, 
\begin{equation}
\label{psi2-corr}
{\overline{\psi^2_l(i) \psi^2_l(j)} - \overline{ \psi^2_l(i)} \
\overline{ \psi^2_l(j)} \over \sigma(\psi^2_l(i)) \sigma(\psi^2_l(j))}
= -{1 \over m-1}\;,
\end{equation}
where $\sigma^2(\psi^2_l(i))$ is the variance of $\psi^2_l(i)$.
This correlator is rather small and in the numerical simulations
exhibits large fluctuations. However, on average it agrees with the
analytical result Eq.~(\ref{psi2-corr}). 

In Sections~\ref{HF-spect} and \ref{HF-wavefunctions}, we have thus
shown that the fluctuation properties of the HF ensemble are very
close to those of RMT: The single--particle HF wave functions obey
the same statistics, and the single--particle HF eigenvalues do, too,
except near the HF gap which plays a special role in the spectrum.
The correlations between filled and empty levels are weakened by the
gap.

\section{Hartree--Fock approach to disordered systems: addition of
  electrons}\label{addition}

In the HF approximation, the many--particle ground--state energy
$E_{\rm HF}(n)$ of $n$ electrons can be written in several ways. In
an arbitrary fixed single--particle basis, we have
\begin{equation}\label{HF-gs}
E_{\rm HF}(n) = \sum_{\alpha \gamma} h^{(0)}_{\alpha\gamma}
\rho_{\gamma\alpha} + \frac{1}{2} \sum_{\alpha\gamma \atop
  \beta\delta} \rho_{\gamma \alpha} v_{\alpha \beta; \gamma
  \delta}\rho_{\delta \beta} \;,
\end{equation}
where $\rho$ is the self--consistent density matrix of
Eq.~(\ref{density-matrix}).

An alternative expression (in terms of the single--particle HF
energies) is
\begin{equation}\label{HF-gs-1}
E_{\rm HF}(n)= \sum_{\alpha = 1}^n \epsilon^{(n)}_{\alpha}  -
\frac{1}{2} \sum_{\alpha\gamma \atop \beta\delta} \rho_{\gamma \alpha}
v_{\alpha \beta; \gamma \delta}\rho_{\delta \beta} \;,
\end{equation}
where the double counting of the interaction terms (in the sum of the
single--particle HF energies) is corrected by the second term on the
r.h.s. of Eq.~(\ref{HF-gs-1}). Both of these expressions simplify when
written in the HF basis [in which $\rho_{\gamma\alpha} = \delta_{\alpha
\gamma} \Theta(n-\alpha)$]
\begin{eqnarray}
\label{peak}
E_{\rm HF}(n)& = & \sum_{\alpha=1}^n h^{(0)}_{\alpha\alpha} + \frac{1}{2}
\sum_{\alpha, \beta=1}^n v_{\alpha \beta;\alpha \beta} \nonumber \\
& = &\sum_{\alpha=1}^n \epsilon^{(n)}_{\alpha} - \frac{1}{2}
\sum_{\alpha,\beta=1}^n v_{\alpha \beta; \alpha \beta} \;.
\end{eqnarray}

\subsection{Peak spacing distribution}\label{peak-spacing}

The position (as given by the gate voltage) of a Coulomb--blockade
peak at low temperature is indicative of the change in the
ground--state energy of a quantum dot due to the addition of an
electron (known as the addition energy). The spacing $\Delta_2$
between successive Coulomb--blockade peaks (for short: the peak
spacing) is then given by the second--order difference of the
ground--state energies versus particle number. In the HF approximation
we have
\begin{equation}\label{delta2}
\Delta_2 \approx E_{\rm HF}(n+1) + E_{\rm HF}(n-1) -2 E_{\rm HF}(n) \;.
\end{equation}
In the CI model, i.e., when only the average interaction is taken into
account, $\Delta_2 = (\epsilon_{n+1} - \epsilon_n) + v_0$, and the
distribution of $\Delta_2$ is a shifted Wigner distribution.
Experimentally, the distribution is closer to a
Gaussian,\cite{sivan96,simmel97,patel98} and this was understood to
be a residual interaction effect.

\begin{figure}[ht]
\vspace{5 mm}
\centerline{\includegraphics[scale=1.0]{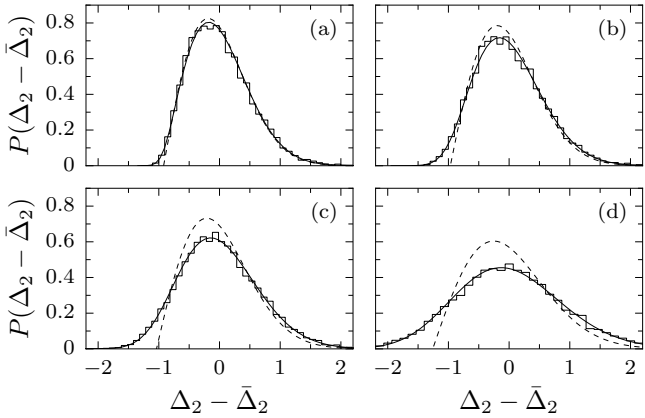}}
\vspace{3 mm}
\caption {Peak spacing distribution $P(\Delta_2 - \bar \Delta_2)$ for
  the same set of $u$--values as in Fig.~\ref{hfgap}.}
\label{dm}
\end{figure}

\begin{figure}[ht]
\vspace{5 mm}
\centerline{\includegraphics[scale=1.0]{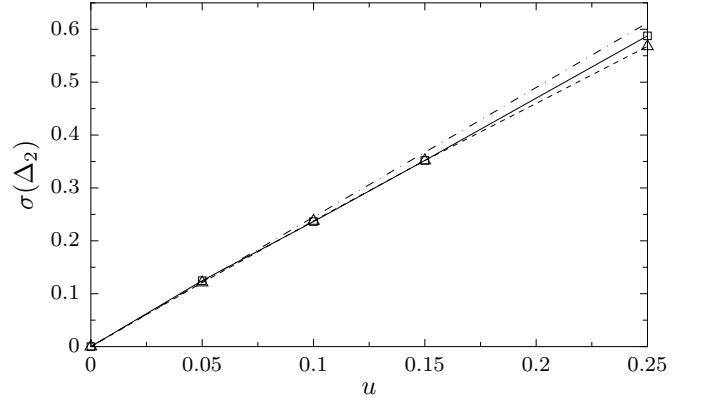}}
\vspace{3 mm}
\caption {The squares depict the width $\sigma$ of a Gaussian the
  convolution of which with a Wigner distribution is fitted to the
  peak spacing distribution $P(\Delta_2 - \bar \Delta_2)$ in
  Fig.~\ref{dm}. Dotted--dashed line, triangles, and dashed line
  are as in Fig.~\ref{hfgapparam}.}
\label{dmparam}
\end{figure}

The histograms in Fig.~\ref{dm} show $P(\Delta_2 - \bar \Delta_2)$ in
the HF approximation for the same cases as in Fig.~\ref{hfgap}. The
solid lines are again convolutions of a Wigner distribution with a
Gaussian of width $\sigma$ for the spacing distribution (this will be
motivated in the following Section). The fitted values of $\sigma$ are
shown versus $u$ as squares in Fig.~\ref{dmparam}. The dotted--dashed
line, triangles, and dashed line are the same as in
Fig.~\ref{hfgapparam}. We note that the average interaction
contributes to $\Delta_2$.

In the spinless case, the average interaction is given by
Eq.~(\ref{average-int}), contributing a constant $v_0$ to $\Delta_2$,
which is canceled out in $\Delta_2-\bar\Delta_2$. Thus the
distribution $P(\Delta_2-\bar\Delta_2)$ is not affected by the average
interaction. This would no longer be true if spin were included: The
average interaction in Eq.~(\ref{average-int-s}) has additional terms
(i.e., an exchange interaction, and, in the orthogonal case, a Cooper
channel interaction), which are not constant for a fixed number of
electrons.

\subsection{Koopmans' approach}

As an electron is added to the quantum dot, the self--consistent
single--particle HF wave functions are expected to change. In
Koopmans' limit, this change is neglected.\cite{koopmans34} This
assumption implies predictions for the peak--spacing distribution
which we now test.

We first calculate the addition energy $E_{\rm HF}(n+1) - E_{\rm
  HF}(n)$ in Koopmans' limit. We use the HF wave functions for $n$
electrons to write expressions for $E_{\rm HF}(n)$ and $E_{\rm
HF}(n+1)$. For $E_{\rm HF}(n)$ we use the exact HF relation $E_{\rm
  HF}(n) = \sum_{\alpha =1}^n h^{(0)}_{\alpha \alpha} + \frac{1}{2}
  \sum_{\alpha \beta =1}^{n} v_{\alpha \beta}$. Using the same
single--particle wave functions, we write for $E_{\rm HF}(n+1)$ the
approximate relation $E_{\rm HF}(n+1) \approx \sum_{\alpha = 1}^{n+1}
  h^{(0)}_{\alpha \alpha} + \frac{1}{2} \sum_{\alpha \beta = 1}^{n+1}
  v_{\alpha \beta}$. We obtain
\begin{equation}\label{add}
E_{\rm HF}(n+1) - E_{\rm HF}(n) \approx h^{(0)}_{n+1,n+1} + 
 \sum_{\beta=1}^{n+1} v_{n+1, \beta} = \epsilon^{(n)}_{n+1}\;.
\end{equation}

A similar expression can be obtained for the $n-1 \to n$ transition,
i.e., $E_{\rm HF}(n) - E_{\rm HF}(n-1) \approx \epsilon^{(n-1)}_{n}$,
where we have used the HF wave functions of $n-1$ electrons. We then
find from Eq.~(\ref{delta2}) the approximate relation\cite{blanter97}
\begin{eqnarray}\label{K1}
\Delta_2 \approx \epsilon^{(n)}_{n+1} - \epsilon^{(n-1)}_{n} & = & 
(\epsilon^{(n)}_{n+1} - \epsilon^{(n-1)}_{n+1}) +
(\epsilon^{(n-1)}_{n+1} - \epsilon^{(n-1)}_{n}) \nonumber \\
&\approx&  v_{n+1,n} + s \;,
\end{eqnarray}
where $s= \epsilon^{(n-1)}_{n+1} - \epsilon^{(n-1)}_{n}$ is the
spacing between the two lowest empty levels for $n-1$ electrons. In
deriving the second approximate relation in Eq.~(\ref{K1}), we have
also used $\epsilon^{(n)}_{n+1} \approx \epsilon^{(n-1)}_{n+1}) +
v_{n+1,n}$, an expression that follows from Koopmans' limit for the
$n-1 \to n$ transition.

An expression analogous to Eq.~(\ref{add}) can be derived from the
$n+1 \to n$ transition, assuming the single--particle HF wave
functions do no change upon the removal of an electron from the dot.
Using the HF wave functions of $n+1$ electrons, we find 
\begin{equation}\label{rem}
E_{\rm HF}(n+1) - E_{\rm HF}(n) \approx \epsilon^{(n+1)}_{n+1} \;.
\end{equation}
Applying a similar relation for the $n \to n-1$ transition, we find
another expression for $\Delta_2$,
\begin{eqnarray}\label{K2}
\Delta_2 \approx \epsilon^{(n+1)}_{n+1} - \epsilon^{(n)}_{n} & = &
(\epsilon^{(n+1)}_{n+1} - \epsilon^{(n+1)}_{n}) +
(\epsilon^{(n+1)}_{n} - \epsilon^{(n)}_{n}) \nonumber \\ &\approx & s + v_{n+1,n} \;,
\end{eqnarray}
where $s= \epsilon^{(n+1)}_{n+1} - \epsilon^{(n+1)}_{n}$ is the
spacing between the two highest filled levels for $n+1$ electrons.

If the HF levels and HF interaction matrix elements are uncorrelated,
Eqs.~(\ref{K1}) and (\ref{K2}) suggest that the distribution of
$\Delta_2$ can be described by a convolution of a Wigner distribution
(with $\Delta$ given by the average spacing of the two lowest empty or
two highest filled levels in the dot, respectively), and a Gaussian
whose width $\sigma$ is given by the standard deviation of the HF
matrix element $v_{n+1,n}$. The results shown in Fig.~\ref{dm} confirm
that the peak spacing distribution is well described by a convolution
of a Wigner distribution with a Gaussian ($\Delta$ is determined from
a mean level density that excludes the gap). Furthermore, for small
values of $u$ ($\alt 0.15$), the width of the Gaussian is well
described by $\sigma(v_{n+1,n})$ (see Fig.~\ref{dmparam}). For larger
values of $u$, we observe some deviations.

\begin{figure}[ht]
\vspace{5 mm}
\centerline{\includegraphics[scale=1.0]{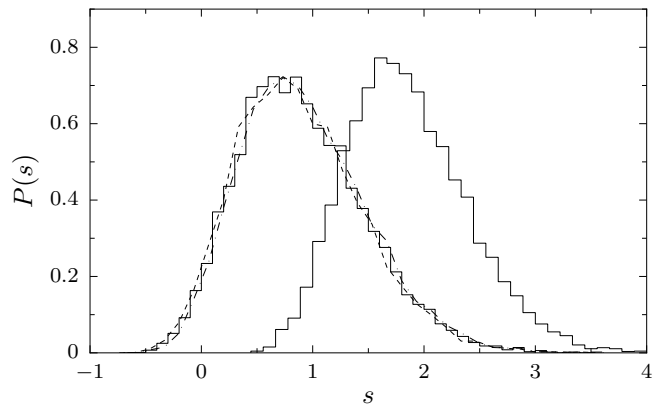}}
\vspace{3 mm}
\caption {The peak spacing distribution $P(\Delta_2)$ (left histogram)
  is compared with $P(\epsilon^{(n+1)}_{n+1} - \epsilon^{(n)}_{n})$
  (dashed line), $P(\epsilon^{(n)}_{n+1} - \epsilon^{(n-1)}_{n})$
  (dotted--dashed line) and the gap distribution (right histogram).
  The ensemble has $u=0.1$.}
\label{hfgapdmcompare}
\end{figure}

In view of Eqs.~(\ref{K1}) and (\ref{K2}), it is interesting to
compare directly the peak spacing distribution $P(\Delta_2)$ with the
distributions  $P(\epsilon^{(n+1)}_{n+1} - \epsilon^{(n)}_{n} )$ and
$P(\epsilon^{(n)}_{n+1} - \epsilon^{(n-1)}_{n})$ using the exact
self--consistent single--particle HF energies (i.e., not using
Koopmans' limit).\cite{walker99}  The results are shown in
Fig.~\ref{hfgapdmcompare} for $u = 0.1$. Both
$P(\epsilon^{(n+1)}_{n+1} - \epsilon^{(n)}_{n} )$ and
$P(\epsilon^{(n)}_{n+1} - \epsilon^{(n-1)}_{n})$ approximate
$P(\Delta_2)$ very well.

If we use Koopmans' limit simultaneously for the addition and removal
of an electron in an $n$--electron dot, i.e., for the transitions $n
\to n+1$ and $n \to n-1$, we obtain the relation $\Delta_2 \approx
\epsilon^{(n)}_{n+1} - \epsilon^{(n)}_{n}$. The quantity
$\epsilon^{(n)}_{n+1} - \epsilon^{(n)}_{n}$ is just the gap in the
$n$--electron dot. This fact motivates the description of the gap
distribution in Fig.~\ref{hfgap} as a convolution of a Wigner
distribution with a Gaussian. The width $\sigma$ of the Gaussian
(found from the gap distribution) is rather close to $\sigma(v_{n+1,
n})$ (see Fig.~\ref{hfgapparam}), although not as close as the value
of $\sigma$ found from the peak--spacing distribution (see
Fig.~\ref{dmparam}).

In Fig.~\ref{hfgapdmcompare} we also compare the gap distribution
$P(\epsilon^{(n)}_{n+1} - \epsilon^{(n)}_{n})$ with the peak--spacing
distribution (without subtracting the average values of the respective
quantities). We observe that the gap distribution is similar in shape
to $P(\Delta_2)$ (both are convolutions of a Wigner distribution with
a Gaussian) but is shifted to the right. This suggests that the
approximation $\Delta_2 \approx \epsilon^{(n)}_{n+1} -
\epsilon^{(n)}_{n}$ does not work as well as Eqs.~(\ref{K1}) and
(\ref{K2}), in particular for the average values.

The qualitative difference in the various approximations for
$\Delta_2$ can be explained as follows. The single--particle HF
energies $\epsilon^{(n)}_{n+1}$ and $\epsilon^{(n-1)}_{n}$ in
Eq.~(\ref{K1}) correspond to empty levels and thus contain both HF
matrix elements connecting empty and filled orbitals. The average
values of these HF matrix elements are similar (see, e.g., in
Fig.~\ref{av} and Eq.~(\ref{HF-element2})) and cancel out when
their differences are taken to find $\Delta_2$. Similarly, the HF
energies $\epsilon^{(n+1)}_{n+1}$ and $\epsilon^{(n)}_{n}$ in
Eq.~(\ref{K2}) both correspond to filled levels. Both contain HF
matrix elements between two filled levels whose averages are again
similar and cancel out upon taking their difference. Because of this
cancellation effect, the approximations in Eqs.~(\ref{K1})and
(\ref{K2}) work better than those in Eqs.~(\ref{add})and (\ref{rem}),
respectively. However, the gap is the difference between an empty
level and a filled level ($\epsilon^{(n)}_{n+1}$ and
$\epsilon^{(n)}_{n}$, respectively). The HF interaction matrix
elements are of the type empty--filled for the empty level and
filled--filled for the filled level. The average values of these
different types of matrix elements are rather different and their
differences contribute to the large average HF gap observed in
Figs.~\ref{hfgapparam} and \ref{hfgapdmcompare}. We note that both
$\Delta_2$ and the gap include a large additional constant $v_0$
(mostly charging  energy) which is not shown in our numerical results
(since we set the average interaction in the original ensemble to
zero).

Another motivation for comparing the peak--spacing distribution with
the above three distributions (using in each case the exact HF levels
for the appropriate number of electrons) is provided by inequalities
for the addition energy that are exact in the HF approximation. The
basic inequalities are~\cite{walker99}
\begin{equation}\label{ineq-add}
\epsilon_{n+1}^{(n+1)} \leq E_{\rm HF}(n+1) - E_{\rm HF}(n)
 \leq \epsilon_{n+1}^{(n)} \;.
\end{equation}
Both inequalities follow from the variational principle for the HF
ground--state energy. To obtain the right inequality, we use the
variational principle for $n+1$ electrons. As a trial wave function we
choose the Slater determinant of the lowest $n+1$ HF wave functions
found in the exact HF solution for $n$ electrons to obtain $E_{\rm
  HF}(n+1) \leq \sum_{\alpha = 1}^{n+1} h^{(0)}_{\alpha \alpha} +
\frac{1}{2} \sum_{\alpha \beta = 1}^{n+1} v_{\alpha \beta}$. Combined
with the exact relation $E_{\rm HF}(n) = \sum_{\alpha =1}^n
h^{(0)}_{\alpha \alpha} + \frac{1}{2} \sum_{\alpha \beta =1}^{n}
v_{\alpha \beta}$, we find the right inequality in (\ref{ineq-add}).
Similarly, the left inequality in (\ref{ineq-add}) is obtained by
using a trial Slater determinant composed of the lowest $n$ orbitals 
found in the exact HF solution for $n+1$ electrons to put an 
upper bound on $E_{\rm HF}(n)$.

Rewriting (\ref{ineq-add}) with $n$ replaced by $n-1$, we obtain HF
inequalities for the ``removal'' energy $E_{\rm HF}(n- 1) - E_{\rm
  HF}(n)$
\begin{equation}\label{ineq-rem}
- \epsilon_{n}^{(n-1)} \leq E_{\rm HF}(n- 1) - E_{\rm HF}(n)
 \leq - \epsilon_{n}^{(n)} \;.
\end{equation}
The peak spacing $\Delta_2$ is obtained by summing the addition and
removal energies in Eqs.~(\ref{ineq-add}) and (\ref{ineq-rem}). We
find that the gap is an upper bound for $\Delta_2$, i.e., 
\begin{equation}
\Delta_2 \leq \epsilon_{n+1}^{(n)} - \epsilon_{n}^{(n)} \;.
\end{equation}
This relation is consistent with the shift to the right of the gap
distribution relative to the peak--spacing distribution (see
Fig.~\ref{hfgapdmcompare}). However, inspecting the inequalities in
(\ref{ineq-add}) and (\ref{ineq-rem}), we conclude that the quantities
$\epsilon^{(n)}_{n+1} - \epsilon^{(n-1)}_{n}$ and
$\epsilon^{(n+1)}_{n+1} - \epsilon^{(n)}_{n} $ are neither upper nor 
lower bounds for $\Delta_2$. As discussed above, they turn out to be
rather good approximations for $\Delta_2$ (for $u \alt 0.15$). 

\subsection{The gap and peak--spacing distributions in the small $u$
  limit}

In Sections~\ref{gap} and \ref{peak-spacing}, we have shown that the
gap and peak--spacing distributions can be approximated as
convolutions of a Wigner distribution (evaluated with the HF
mean--level spacing) and a Gaussian whose width $\sigma$ is close to
$\sigma(v_{n+1,n})$ (for the ``local'' interaction ensemble
$\sigma(v_{n+1,n}) \approx \sqrt{6} u$). On might have chosen another
approach: One may attempt to use Eq.~(\ref{HF-gap}) directly to infer
the distribution of, e.g., the gap in the limit $u \to 0$. The various
diagonal HF interaction matrix elements in Eq.~(\ref{HF-gap}) are
approximately uncorrelated Gaussian variables with variance
$\sigma^2(v_{n+1,n}) \approx 6 u^2$ each. Furthermore, to first order
in the interaction, we have $h^{(0)}_{\alpha, \alpha} \approx
\epsilon_\alpha$ and $v_{\alpha, \beta} \approx v^{(0)}_{\alpha,
  \beta}$. Within RMT the quantities $\epsilon_\alpha$ and $
v^{(0)}_{\alpha, \beta}$ are uncorrelated. One might then conclude
from Eq.~(\ref{HF-gap}) that, in the limit $u \to 0$, the gap
distribution is a convolution of a Wigner distribution (evaluated with
the GOE value of $\Delta$ without interaction) and a Gaussian whose
width is $\approx \sigma = \sqrt{2n-1}\; \sigma(v^{(0)}_{n+1,n})
\sqrt{6} u$ [there are $2n-1$ interaction matrix elements in
Eq.~(\ref{HF-gap})].

\begin{figure}[ht]
\vspace{5 mm}
\centerline{\includegraphics[scale=0.43]{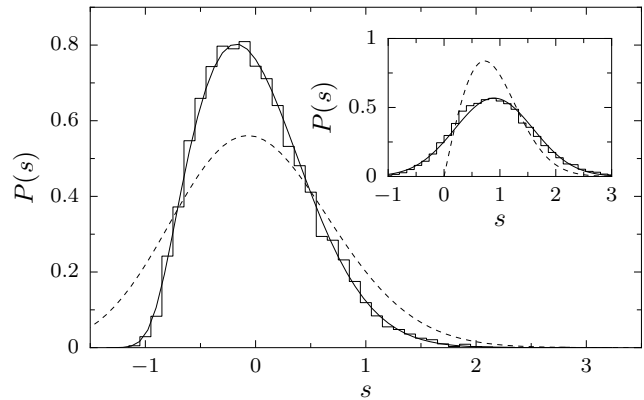}}
\vspace{3 mm}
\caption {The peak spacing distribution $P(\Delta_2)$ (histogram)
  for the ``local'' interaction ensemble with $u=0.05$ is compared with
  a convolution of a Wigner distribution with a Gaussian whose width is
  $\sigma(v^{(0)}_{n+1,n})$ (solid line). The dashed line describes a
  convolution of a Wigner distribution and a Gaussian with a width of
  $\sqrt{2n-1} \;\sigma(v^{(0)}_{n+1,n})$. Inset: the distribution of
  $h^{(0)}_{n+1,n+1} - h^{(0)}_{n,n}$ (histogram) is approximately
  fitted by a Gaussian (solid line). The dashed line is the Wigner
  distribution of $\epsilon_{n+1} -\epsilon_{n}$. The number of
  electrons is $n = 10$.}
\label{hfgapcompare}
\end{figure}

When compared with our previous results, the width $\sigma$ of the
Gaussian would then be enhanced by the large factor $\sqrt{2n-1}$.
This increase is not compensated by the small decrease in the width of
the Wigner distribution (the GOE value of $\Delta$ is a little smaller
than the HF value $\Delta_{\rm HF}$). The result is a distribution
which is much wider than the one found in the numerical simulation.
We show in Fig.~\ref{hfgapcompare} the gap distribution $P(\Delta_2)$
(histogram) for a small value of $u$ ($u=0.05$). The solid line,
predicted by Eq.~(\ref{K1}) or Eq.~(\ref{K2}), is a convolution of a
Wigner distribution (with the appropriate $\Delta$ at $u=0.05$) and a
Gaussian with a width of $\sigma= \sigma(v^{(0)}_{n+1,n})$. As
expected, the agreement is very good. However, the distribution that
one infers using Eq.~(\ref{HF-gap}) in the limit $u \to 0$ (dashed
line) is much broader. The discrepancy is due to the fact that even
for very small values of $u$, the difference $h^{(0)}_{n+1,n+1} -
h^{(0)}_{n,n}$ is strongly correlated with the HF matrix elements.
Hence, the assumption made in using Eq.~(\ref{HF-gap}) is not
correct. The inset of Fig.~\ref{hfgapcompare} shows the distribution
of $h^{(0)}_{n+1,n+1} - h^{(0)}_{n,n}$ (histograms). This distribution
can be approximately fitted by a Gaussian (solid line), but is
distinctly different from the Wigner distribution of $\epsilon_{n+1}
-\epsilon_{n}$ (dashed line).

\subsection{Peak--height distribution}

To measure the conductance, it is necessary to couple the dot to
external leads. For weak coupling (i.e., for an almost isolated dot)
and at low temperature ($T \ll \Delta$), the conductance peak height
can be expressed in terms of the ground--state wave functions of the
dot with $n$ and $n+1$ electrons, and we can use the formalism
developed here to study the statistical properties of the conductance. 
In terms of the rates $\Gamma^l$ and $\Gamma^r$ for an electron to
tunnel into the dot from the left and right leads, respectively, the
peak height is given by
\begin{equation}
\label{peak-height}
G \propto {e^2\over \hbar k T} {\Gamma^l \Gamma^r \over \Gamma^l +
  \Gamma^r}\;.
\end{equation}
The rates are squares of partial--width amplitudes, $\Gamma^{l,r} =
|\gamma^{l,r}|^2$. In the CI model, the $\gamma$'s are proportional to
the projection of a single--particle eigenfunction (occupied by the
electron that tunnels into the dot) on the respective lead (or point
contact ${\bf r}$). For symmetric leads that are separated by a distance large compared with the Fermi wavelength, the amplitudes $\gamma^l$ and $\gamma^r$ are
uncorrelated Gaussian random variables with zero mean value and a
common second moment. This leads to the well--known predictions for
the peak--height distributions.\cite{JSA92} 

A study of the peak height statistics in a small dot described by an Anderson model plus Coulomb interactions suggested that the peak--height distribution is only weakly affected by residual interactions.\cite{berkovits98a}  To find out whether this result is generic, we use in the following the HF approximation to study the peak-height distribution within the framework of the induced two--body ensembles. 

In the presence of interactions (beyond charging energy), we have
\begin{equation}
\gamma \propto \langle \Phi(n+1) | \hat\psi^\dagger({\bf r})| \Phi(n)
\rangle \;,
\end{equation}
where $\hat \psi^\dagger({\bf r})$ creates an electron at the point
contact ${\bf r}$, and $\Phi(n)$ is the ground--state wave function
of the dot with $n$ electrons. In the HF approximation, $\Phi(n)$ is
a Slater determinant of the lowest $n$ single--particle HF wave
functions $\psi_1,\ldots,\psi_n$.

In Koopmans' limit, the single--particle HF wave functions do not
change with $n$. Expanding $\hat \psi^\dagger({\bf r}) = \sum_\lambda
\psi_\lambda^*({\bf r}) a^\dagger_\lambda$ where $a_\lambda^\dagger$
creates an electron in the single--particle HF state $\psi_\lambda$
of $n$ electrons, we find
\begin{equation}\label{gamma-Koopmans}
\gamma \propto \psi_{n+1}^*({\bf r}) \;
\end{equation}
for the $n \to n+1$ transition. This expression is similar to the
expression found for the CI model except that the single--particle
HF wave function replaces the non--interacting wave function. 

We have shown in Section~\ref{HF-wavefunctions} that the statistics of
the single--particle HF wave functions are the same as in standard
RMT. In particular, the distribution of the components $\psi_{n+1}(j)$
of the HF wave function $\psi_{n+1}$ in a fixed basis $|j\rangle$ is
given by Eq.~(\ref{dist-vector}). For a chaotic ballistic dot such a
fixed basis $\chi_j({\bf r})$ is given by the free particle states
with energy $\epsilon = \hbar^2 k^2/2m$, e.g., circular waves
$\chi_j({\bf r}) \propto J_j(k r) e^{ij\theta}$ ($ j=0, \pm1, \pm2
\ldots$) where $J_j$ are Bessel functions of the first kind. A
single--particle eigenfunction of the non--interacting dot can be
expanded $\psi_\alpha({\bf r}) = \sum_j \psi_\alpha(j) \chi_j({\bf
  r})$ and, for a chaotic dot, the components $\psi_\alpha(j)$ follow
RMT statistics. Expanding the HF wave function $\psi_{n+1}$ in the
same basis $\psi_{n+1}({\bf r}) = \sum_j \psi_l(j) \chi_j({\bf r})$,
we can view $\psi_{n+1}({\bf r})$ as a projection of $\psi_{n+1}$ on a
fixed vector whose components are $\chi_j({\bf r})$. Because of the
orthogonal (unitary) invariance of the HF ensemble, such a projection
has the same distribution as the distribution~(\ref{dist-component})
of an eigenvector component. Thus the distribution of $\Gamma =
|\gamma|^2$ is given by (\ref{y-dist}) and approaches a Porter--Thomas
distribution for large $m$. Using $\overline{\psi^*_{n+1}(i)
  \psi_{n+1}(j)} =\delta_{ij}/m$, we also find (see, for example,
Section V.B in Ref.~\onlinecite{rmp00}) 
\begin{equation}
{\overline{{\gamma^l}^* \gamma^r} \over
  \sigma(\gamma^l)\sigma(\gamma^r)} = J_0(k|{\bf r_l} - {\bf r_r}|)
\;,
\end{equation}
where $\sigma(\gamma)$ is the standard deviation of the partial
amplitude $\gamma$. Thus the correlation between the left and right
partial width amplitudes decays in magnitude as $\sim (k |{\bf r}_l -
{\bf r}_r|)^{-1/2}$, and can be ignored when the distance between the
left and right leads is large compared with the Fermi wavelength.

Since $\Gamma^l$ and $\Gamma^r$ are uncorrelated and each follows a
Porter--Thomas distribution, we conclude (in Koopmans' limit) that the
peak--height distribution for the orthogonal (unitary) many--body
ensembles of Section~\ref{ran} is the same as the distribution derived
from standard RMT.

Beyond Koopmans' limit, the single--particle HF wave functions depend
on the number of electrons on the dot and are accordingly denoted by
$\psi^{(n)}_\lambda$. Expanding in the HF basis of $n$ electrons,
$\hat\psi^\dagger({\bf r})= \sum_\lambda \psi_\lambda^{(n)*}({\bf r})
a_\lambda^\dagger$ (where $a_\lambda^\dagger$ creates an electron in
$\psi^{(n)}_\lambda$), we find
\begin{equation}\label{HF-gamma}
\!\langle \Phi_{\rm HF}(n+1) | \hat\psi^\dagger({\bf r})|\Phi_{\rm
  HF}(n)\rangle = \!(-)^n \!\!\!\!\sum_{\lambda = n+1}^m \!\!
  {\psi_\lambda^{(n)}}^*({\bf r}) \det A^{(\lambda)}, 
\end{equation}
where $A^{(\lambda})$ is an $(n+1)\times(n+1)$ matrix defined by
$A^{(\lambda)}_{\alpha\beta} = \langle \psi^{(n+1)}_\alpha |
\psi^{(n)}_\beta \rangle$ and $A^{(\lambda)}_{\alpha,n+1} = \langle
\psi^{(n+1)}_\alpha | \psi^{(n)}_\lambda \rangle$ for $\alpha=1,
\ldots, n+1$; $\beta=1,\ldots,n$.

An equivalent expression for $\gamma$ can be derived by rewriting 
$\gamma \propto \langle \Phi_{\rm HF}(n) | \hat\psi({\bf r})|\Phi_{\rm
  HF}(n+1)\rangle^*$ and expanding in the HF basis of $n+1$ electrons
$\hat \psi({\bf r}) = \sum_\lambda \psi^{(n+1)}_\lambda({\bf r})
a_\lambda$ (where now $a_\lambda$ annihilates an electron in
$\psi^{(n+1)}_\lambda$). We find
\begin{equation}
\label{HF-gamma1}
\!\!\!\langle \Phi_{\rm HF}(n+1) | \hat\psi^\dagger({\bf r})
| \Phi_{\rm HF}(n)\rangle = \!\! \sum_{\lambda =
1}^{n + 1}\! (-)^{\lambda-1} \psi_\lambda^{(n + 1)*}({\bf r})
\det B^{(\lambda)},
\end{equation}
where $B^{(\lambda)}$ is an $n \times n$ matrix defined by
$B^{(\lambda)}_{\alpha \beta} = \langle \psi^{(n+1)}_\alpha |
\psi^{(n)}_\beta \rangle$ for $\alpha = 1, \ldots, \lambda - 1,
\lambda + 1, \ldots, n+1$ and $\beta = 1, \ldots, n$. 

\begin{figure}[ht]
\vspace{5 mm}
\centerline{\includegraphics[scale=1.0]{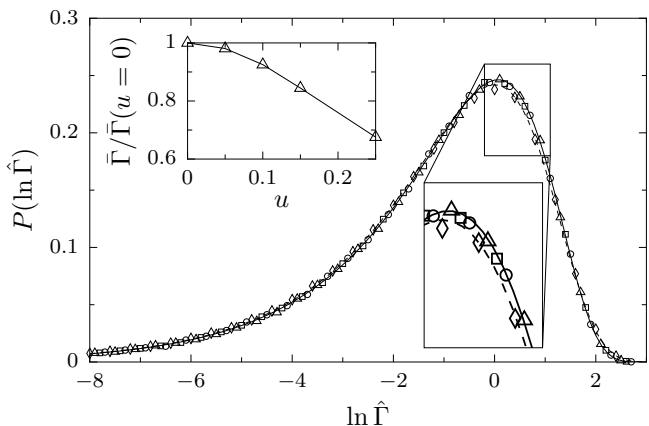}}
\vspace{3 mm}
\caption {The distribution  $P(\ln \hat\Gamma)$ versus $\ln
  \hat\Gamma$ where $\hat\Gamma$ is the renormalized partial width in
  the HF approximation. Shown are results for $u = 0.05$ (triangles),
  $0.1$ (squares), $0.15$ (circles) and $0.25$ (diamonds). We use the
  ensemble~(\ref{corr-local-a}) with $m = 30$ and $n=10$. The solid
  line is the distribution~(\ref{y-dist}) and the dashed line is the
  Porter-Thomas distribution. We have magnified a section of the graph
  to show more details. Inset: $\bar\Gamma / \bar \Gamma(u = 0)$
  versus $u$.}
\label{Gamma}
\end{figure}

Taking for $\psi^{(n)}_\lambda({\bf r})$ a projection of the HF wave
function $\psi^{(n)}_\lambda$ on a fixed vector, we have used 
Eq.~(\ref{HF-gamma}) to calculate the distribution of the renormalized
partial width $\hat \Gamma= \Gamma/\bar \Gamma$ in the HF approach
(alternatively, we can use Eq.~(\ref{HF-gamma1}) with
$\psi^{(n+1)}_\lambda({\bf r})$ given by a projection of the HF wave
function $\psi^{(n + 1)}_\lambda$). Fig.~\ref{Gamma} shows (symbols)
the distribution $P(\ln \hat \Gamma)$ versus $\ln \hat \Gamma$ for $u
= 0.05, 0.1, 0.15$ and $0.25$. The solid line is the
distribution~(\ref{y-dist}) for $m = 30$ and the dashed line is the
Porter--Thomas distribution. For $u = 0$, only the first (last) term
in Eq.~(\ref{HF-gamma}) (Eq.~(\ref{HF-gamma1})) differs from zero, and
the distribution is given by Eq.~(\ref{y-dist}). As $u$ increases,
more terms contribute and the distribution of $\hat \Gamma$ gets
closer to the Porter--Thomas distribution (this is plausible because
of a central--limit theorem). The inset of Fig.~\ref{Gamma} shows the
average value of $\Gamma$ in units of the average value of $\Gamma$ in
the CI model. This average value decreases from $1$ as $u$
increases. We have also calculated correlations of partial widths
(e.g., $(\overline{\Gamma^l \Gamma^r} - \bar\Gamma^l \bar\Gamma^r) /
[\sigma(\Gamma^r) \sigma(\Gamma^l)]$), and found them to be generally
weaker than predicted by Eq.~(\ref{psi2-corr}).

We conclude that in the HF approach, the partial--width distribution
is not affected by the fluctuations of the residual interaction and
(in the limit of large $m$) remains Porter--Thomas like. Consequently,
the peak--height distribution is similar to the distribution predicted
from standard RMT. The residual interaction only affects the average
values of the partial widths.

\section{Summary and conclusion}

In this paper we have discussed a generic approach towards the
understanding of the statistical properties of an (almost) closed
diffusive or chaotic quantum dot in the limit of large but finite
Thouless conductance $g$.

The randomness of the single--particle Hamiltonian of a closed
diffusive or chaotic dot induces randomness into the two--body
interaction matrix elements when these are expressed in the
eigenbasis of the single--particle Hamiltonian. In the first part
of this paper we have classified the resulting induced two--body
ensembles. These depend on both, the underlying space--time
symmetries of the dot, and the symmetries of the interaction
matrix elements under permutations of the single--particle basis.
Our classification applies to the non--antisymmetrized matrix 
elements and therefore holds in the presence of spin degrees 
of freedom. The ensembles for the spinless case follow directly 
by antisymmetrizing the interaction matrix elements. We have
ignored spin--orbit scattering in the dot, so the resulting
two--body ensembles have either orthogonal or unitary symmetry
depending on whether time--reversal symmetry is conserved or
broken. Aside from the symmetries, the ensembles are
characterized by three parameters: The mean level spacing
$\Delta$ due to the random one--body Hamiltonian, the parameter
$u$ which measures the fluctuations of the two--body interaction
matrix elements, and the number $n$ of electrons on the dot.

The presence of two--body interactions poses difficulties in
treating the many--body system. Therefore, in the second part
of this work we have used the HF approximation to study generic
interaction effects in an (almost) closed dot. The HF
approximation is not only tractable, but also allows us to
use a single--particle formulation, optimized to include
interaction effects.

Applying the HF approximation to the two--body ensembles, we
have developed a generic statistical HF approach. In this
approach we have solved the HF equations for a large number
of realizations of the ensemble, thereby generating a one--body
ensemble of single--particle HF energies and wave functions. We
have studied the statistical properties of this induced HF 
one--body ensemble, distinguishing the lowest $n$ filled levels
from the remaining empty ones. For both the filled and the
empty levels separately, the nearest--neighbor level spacing
follows the Wigner distribution. This is not true for the HF
gap separating the filled and empty levels. The gap distribution
has the form of a convolution of a Wigner distribution with a
Gaussian whose width is proportional to $u$, while the average
value of the gap increases quadratically with $u$ (in leading
order). By studying a suitable spacing correlator, we have found
that the gap weakens the correlation between filled and empty
levels, but similar correlators within the filled or empty levels
separately continue to follow RMT. 
We have shown that the single--particle HF ensemble satisfies
orthogonal (unitary) invariance in the presence (absence) of
time-reversal symmetry. Consequently the HF wave function
 components (in a fixed basis) satisfy RMT statistics. 
We have also studied the statistics of the interaction
matrix elements in the HF basis (which reflect the statistical
properties of the HF wave functions). The distributions of the
HF matrix elements are Gaussian and, thus, can be characterized
by their first two moments. The average of a diagonal interaction
matrix element acquires a $u^2$ correction. The value of that
correction depends on whether the corresponding single--particle
orbitals are both filled, both empty, or whether one is filled,
the other, empty. The variances, on the other hand, remain close
to their values in the non--interacting basis.

The HF ensemble can be used to study the statistical properties
of various observables. In particular, we have studied the generic
properties of the peak--spacing distribution and of the 
peak--height distribution. A simple interpretation of the results
is provided in Koopmans' limit where it is assumed that the HF
wave functions do not change upon the addition of an electron to
the dot. The peak--spacing distribution is well approximated by a
convolution of a Wigner distribution with a Gaussian whose width
is the standard deviation of a diagonal interaction matrix element.
The peak--height distribution, on the other hand, is found to be
insensitive to the residual interactions.

We have confined ourselves to the HF approximation for spinless
electrons. The presence of spin leads to additional technical
difficulties in a mean--field  approach. In particular, only
the $z$--component of total spin (but not total spin itself) is
conserved. Because of this restriction in our calculations, it
is not possible to make a quantitative comparison between our
results and experiment. In the presence of spin, the
single--particle levels are doubly degenerate. Therefore, the
peak--spacing distribution is expected to be bimodal. The
exchange interaction associated with the spin degrees of
freedom explains the finite--temperature suppression of the
width of the peak--spacing distribution,\cite{alhassid03}  but at
low temperatures the distribution is expected to remain bimodal.
No such bimodality is observed in the experiments, however. This
fact is related to the fluctuating part of the interaction (which
is of order $u \propto \Delta/g$). We have seen that in the
spinless case, these fluctuations affect the shape of the
peak--spacing distribution and lead to a distribution which is
intermediate between a Wigner distribution and a Gaussian
distribution, in qualitative agreement with experiments.  It
would be interesting to study the corresponding finite--$g$
effects in the presence of spin, using the HF approximation
within the induced two--body ensembles discussed in
Section~\ref{ran}. This should lead to a generic peak--spacing
distribution that can be directly compared with the experimental
data. As for the peak--height distribution, RMT plus exchange
interaction (without the fluctuating part of the interaction)
were found to provide a quantitative description of the observed
features at low temperatures.\cite{alhassid03,usaj03} We have
shown here that in the absence of spin, the peak--height
distribution is insensitive to the fluctuating residual
interaction. If this conclusion continues to hold in the
presence of spin, it will fully explain the observed
low--temperature peak--height distribution. A definitive answer
could be provided using the induced two-body ensembles of
Section~\ref{ran} that include spin. 

\section*{Acknowledgments}

We thank Y. Gefen, Ph. Jacquod, L. Kaplan and A.D. Mirlin for useful 
discussions. This work was supported in part by the U.S. DOE grant 
No.\ DE-FG-0291-ER-40608.  Y.A. would like to acknowledge
support by a von Humboldt Senior Scientist Award and the
hospitality of the Max-Planck-Institut f\"ur Kernphysik
at Heidelberg where part of this work was completed. A.W.
acknowledges support by the SFB 484.

\end{document}